\documentclass[aps,prd,twocolumn,preprintnumbers,superscriptaddress,floatfix]{revtex4}

\usepackage{multirow, graphicx,amssymb,url,mathrsfs,amsmath}
\usepackage{eucal,wrapfig,boxedminipage,setspace,subfigure}
\usepackage{amsxtra,amstext,latexsym,dsfont}

\def\IR{{\hbox{{\rm I}\kern-.2em\hbox{\rm R}}}}
\def\IB{{\hbox{{\rm I}\kern-.2em\hbox{\rm B}}}}
\def\IN{{\hbox{{\rm I}\kern-.2em\hbox{\rm N}}}}
\def\IC{\,\,{\hbox{{\rm I}\kern-.59em\hbox{\bf C}}}}
\def\IZ{{\hbox{{\rm Z}\kern-.4em\hbox{\rm Z}}}}
\def\IP{{\hbox{{\rm I}\kern-.2em\hbox{\rm P}}}}
\def\IH{{\hbox{{\rm I}\kern-.4em\hbox{\rm H}}}}
\def\ID{{\hbox{{\rm I}\kern-.2em\hbox{\rm D}}}}

\newcommand{\beq}{\begin{equation}}
\newcommand{\eeq}{\end{equation}}
\newcommand{\bea}{\begin{eqnarray}}
\newcommand{\eea}{\end{eqnarray}}

\begin{document}

\voffset 1cm

\newcommand\sect[1]{\emph{#1}---}

\title{Holograms of a Dynamical Top Quark}

\author{Will Clemens, Nick Evans, \& Marc Scott}
\affiliation{ STAG Research Centre \&  Physics and Astronomy, University of
Southampton, Southampton, SO17 1BJ, UK}

\begin{abstract}

\noindent We present holographic desciptions of dynamical electroweak symmetry breaking models that incorporate the top mass generation mechanism. The models allow computation of the spectrum in the presence of large anomalous dimensions due to walking and strong NJL interactions. 
Technicolour and QCD dynamics are described by the bottom-up Dynamic AdS/QCD model for arbitrary gauge groups and numbers of quark flavours. An
assumption about the running of the anomalous dimension of the quark bilinear operator is input and the model then predicts the spectrum and
decay constants for the mesons. We add NJL interactions responsible for flavour physics from extended tecnhicolour, top-colour etc  using 
Witten's multi-trace prescription. We show the key behaviours of a top condensation model can be reproduced. We study generation of the top
mass in (walking) one doublet and one family technicolour models and with strong ETC interactions. The models clearly reveal the tensions between the large top mass and precision data for $\delta \rho$. The neccessary  tunings needed to generate a model compatible with precision constraints are simply demonstrated. 

\end{abstract}

\maketitle

\newpage
\section{Introduction}

Technicolour \cite{Weinberg:1975gm,Susskind:1978ms,Farhi:1980xs,Hill:2002ap} remains an appealing paradigm for breaking electroweak symmetry since it mirrors the symmetry breaking mechansim in QCD and superconductors. It has long faced  a variety of attacks from flavour changing neutral current data \cite{Eichten:1979ah,Dimopoulos:1979es}, precision electroweak data \cite{Peskin:1990zt} and now the discovery of a very fundamental looking higgs state
\cite{Aad:2012tfa,Chatrchyan:2012xdj}. There still perhaps seems a small hope that these issues can be dodged by suitable tuning in the parameter space of the collection of strongly coupled gauge theories. In particular walking theories \cite{Holdom:1981rm}, in which there is a large anomalous dimension for the quark bilinear over a large energy range, might raise the flavour scale, lower the electroweak S parameter \cite{Sundrum:1991rf}, and even generate a light technidilaton type state \cite{Yamawaki:1985zg,Bando:1986bg,Miransky:1996pd,Appelquist:1998xf,Hong:2004td,Dietrich:2005jn}.

The discovery of the top quark 23 years ago \cite{Abe:1995hr} with its very large mass presented the toughest challenge. If one naively uses extended technicolour (ETC) \cite{Eichten:1979ah,Dimopoulos:1979es} interactions to generate the top mass then one expects 
\begin{equation} m_t \simeq  {g^2 \langle \bar{Q} Q \rangle \over \Lambda^2} \simeq {g^2 (4 \pi v^3) \over \Lambda^2} \label{simpletop}\end{equation}
where $Q$ are techni-quark fields, $v$ is the electroweak scale and $\Lambda$ the mass scale of the new interactions generating the top mass. Naturally, with the ETC coupling $g\simeq 1$, $\Lambda$ should be at or below the 1 TeV scale. When one tried to include the isospin violating physics neeeded to generate the top bottom mass splitting at such a low scale deviations in the electroweak precision $\delta \rho$ or T  parameter were of order 100 rather than 0.1 \cite{Chivukula:1995dc,Appelquist:1996kp}! This issue is so confounding that most more recent work on technicolour has concentrated on the core electroweak breaking dynamics and put aside completely the flavour generation mechanism - the top remains the elephant in the room!

Two possible resolutions of the top problem have been suggested. The first is that walking dynamics might enhance the techni-quark condensate and raise $\Lambda$. Twenty years ago gap equation \cite{Appelquist:1986an,Appelquist:1987fc,Appelquist:1989ps,King:1989gq}and Pagel Stokar type formulae \cite{Pagels:1979hd} were the state of the art for addressing this issue but it was hard to generate a sufficient, needed rise in the tail of the techni-quark self energy to raise $\Lambda$ enough \cite{Chivukula:1995dc,Appelquist:1996kp, Appelquist:1989ps}. The second idea was essentially top condensation \cite{Miransky:1988xi,Miransky:1989ds,Nambu:1989jt,Bardeen:1989ds}; additional strong interactions of the top at high scale generated NJL operators that by themselves generated a top condensate and the top mass independently of the technicolour sector which still performed the majority of the work of breaking electroweak symmetry. A mix of these ideas and the possibility that the ETC interactions were also strongly interacting seemed possible but it was hard to construct a computational framework that seemed in anyway reliable. 

In the intervening twenty years a new method for computation in strongly coupled gauge theories has emerged from string theory, holography\cite{Maldacena:1997re}. Holography provides a rigorous method of computation in a selection of strongly coupled gauge theories lying near ${\cal N}=4$ gauge theory. Amongst these theories are those with quarks that display chiral symmetry breaking \cite{Babington:2003vm,Kruczenski:2003uq,Filev:2007gb,Sakai:2004cn}. Some aspects of the meson spectrum are predicted in these models \cite{mesons}. At least in the quenched (probe \cite{Karch:2002sh}) limit the key ingredient to determine the spectrum is the running anomalous dimension of the quark bilinear ($\bar{q}q$)\cite{Jarvinen:2011qe,Kutasov:2011fr,Alvares:2012kr}. Embracing that observation it is possible to construct holographic models of a wider class of gauge theories including those with SU($N_c$) gauge group and $N_f$ quark flavours \cite{Jarvinen:2011qe,Jarvinen:2009fe,Alho:2013dka,Evans:2013vca}. We will use the simple Dynamic AdS/QCD model \cite{Alho:2013dka} in which a plausible guess for the running anomalous dimension is input by hand. Here we will be led by the two loop perturbative running in the gauge theory. This of course is not to be trusted in the non-perturbative regime but these ansatz provide a set of runnings that include a rising IR fixed point value of $\gamma$ as $N_f$ decreases and give candidates for a conformal window and walking theories \cite{Appelquist:1998rb,Dietrich:2006cm}. The predictions for the QCD ($N_c=3,$ $N_f=2$) spectrum lie reasonably close to observation at the 20$\%$ level \cite{Clemens:2017udk}. It is worth stressing that these successes are inspite of a rather brutal truncation of the operators assumed to participate in the dynamics and neglect the expected more stringy aspects of a true description of the physics. Holography is particularly well suited to the study of walking dynamics because the running anomalous dimension is the key input. The expected increase in the quark condensate and a light higgs-like $\sigma$ have been observed in the model in the walking regime \cite{Alho:2013dka} (the lightness of this state has been disputed in the alternative holographic model of \cite{Jarvinen:2011qe,Jarvinen:2009fe} where deep IR conformal symmetry breaking raises the state's mass but other states seem to behave similarly in the different models). One can hope as one moves away to theories with walking behaviour the model will continue to make sensible predictions of the spectrum.

Recently it has been understood how to use Witten's double trace prescription \cite{Witten:2001ua} to include Nambu-Jona Lasinio (NJL) \cite{Nambu:1961tp} four fermion operators in holography and reproduce the usual NJL chiral symmetry breaking behaviour if the coupling lies above some critical value \cite{Evans:2016yas}. The base model, before the introduction of the NJL interaction, generates an effective potential for the model against quark mass (the holographic model computes this by evaluating its action on the vacuum solutions). The Witten prescription includes the NJL operator as a classical piece in the effective potential evaluated on the solution at the cut off $\Lambda$ so
\begin{equation} \Delta V_{Eff} = {g^2 \over \Lambda^2} \bar{Q}_L Q_R \bar{Q}_R Q_L + h.c. = {\Lambda^2 \over g^2} m_Q^2 \end{equation}
where we have used $m_Q = g^2/\Lambda^2 \langle \bar{Q} Q \rangle$. This is formally appropriate in a large $N$ limit. With the NJL term present one allows $m_Q$ to become  dynamically determined and the resulting potential can generate a non-zero mass if $g$ is large enough. This will be the key tool that will enable us to include ETC flavour interactions into the Dynamic AdS/QCD model of technicolour dynamics.

We stress that these holographic models are not first principle computations but they are sensibly motivated descriptions of the dynamics that include the running of the anomalous dimension more directly than other approximations. They allow the construction of, and simple computation in, a full system of the interactions of these models. It's rather pleasing to be able to construct these models within the new holographic formalism from a purely theoretical stand point, even if LHC data is rather constraining hopes for physics beyond the Standard Model. 

We will first review Dynamic AdS/QCD \cite{Alho:2013dka} which we use to describe strongly coupled gauge theories. We will then review the multi-trace prescription for NJL operators \cite{Evans:2016yas,Clemens:2017udk}. Armed with these tools we will first present a hologram of a top condensation model.  We show the critical behaviour for chiral symmetry breaking and the fine tuning needed to achieve $m_t \ll \Lambda$. Holography should be trusted where strong interactions are dominating the dynamics. In top condensation the NJL operator is strong in a regime where all other interactions are weak and the holographic description of the quarks is less secure. AdS/QCD models pass muster in the weakly coupled regime because they contain a memory of N=4 SYM theory which, like the perturbative gauge theory, is near conformal and protects the anomalous dimensions of the operators considered to their perturbative values. In fact the memory of supersymmetry means that the effective potential is flat with quark mass in the absence of running in our holographic model - the expected fermion loop contributions to the effective potential are absent (they have cancelled aginst the squark contribution in the origin theory). When running is introduced, supersymmetry is broken and an effective potential that falls to large quark mass develops allowing the behaviour we have described. The effective potential is dominated by ``cracked egg'' diagrams where gluons are exchanged across the fermion loop. Given this distinction from the basic NJL description of top condensation one does not realize exactly the same critical coupling but all the characteristic behaviours are present. There is also a phenomenological parameter, $\kappa$ (a 5d gauge coupling) which is unfixed in the model and determines $f_\pi$ for a given $m_t$ - for order one values of $\kappa$ the top mass can not generate sufficient $f_\pi$ to explain the electroweak symmetry breaking vev as one expects. 

Our second model is a one electroweak doublet extended technicolour model. Dynamic AdS/QCD allows us to study an SU($N_{TC}$) gauge group with varying number of flavours, $N_f$. Our input in each case is the running anomalous dimension of the quark bilinear, $\gamma$, taken from the two loop perturbative running of the technicolour coupling $\alpha_{TC}$.  The IR fixed point in this approximation crosses through the point $\gamma=1$ where chiral symmetry breaking is triggered for $N_f \simeq 4 N_c$ (the ``edge'' of the conformal window \cite{Dietrich:2006cm}). In the gravity dual this transition corresponds to where the Breitenlohner Freedman (BF) \cite{Breitenlohner:1982jf} bound is violated in the IR by the running mass of the scalar dual to the quark condensate. We will study the $N_{TC}=3$ case and vary $N_f$. At higher $N_{TC}$ one can sample very similar running profiles with less discrete jumps but the $N_{TC}=3$ case suffices to show the main features. For higher $N_{TC}$ similar examples can be found by appropriate choices of $N_f$. 

Here we assume extra techni-quarks beyond the single electroweak doublet (contributing $N_f=2$) are electroweak singlets which allows us to impose walking behaviours for the running on a minimal electroweak sector. These models are perhaps most likely to be compatible with the electroweak S parameter \cite{Peskin:1990zt}. The S parameter essentially counts electroweak doublets and perturbatively a doublet contributes $1 / 6 \pi \simeq 0.05$  to be compared with an experimental upper limit of 0.3. QCD-like strong dynamics are known to increase this contribution by a factor of 2 or more so with $N_{TC}=3$ copies of a single doublet the bound is close to saturation.  It is possible walking dynamics alleviates this isssue \cite{Sundrum:1991rf}. This drop in S as one approaches the edge of the conformal window can be modelled in the Dynamic AdS/QCD model by allowing the parameter $\kappa$ of the model to fall to zero as $N_f \rightarrow 4 N_c$ \cite{Alho:2013dka}. The contribution to S in Dynamic AdS/QCD can be found in Fig 10 of \cite{Alho:2013dka} - we will not address this issue further here.  The need for a low S motivates our restriction to $N_{TC}=3$ also though. 

In this model, since technicolour is strong (even out to the ETC scale in the walking cases), the cracked egg diagram domination of the effective potential is more appropriate and the holographic description of the NJL interaction hopefully sensible.  We put in the four fermion operators of a classic ETC unification to generate the top mass - they link the top to the tecnhi-U quark but not the techni-D quark. We begin by finding solutions for the NJL and TC couplings that generate some given top mass whilst correctly generating the electroweak scale $f_\pi= 246$ GeV. Generically there are two solutions. One matches to the usual weakly coupled ETC regime - for low top masses the technicolour dynamics dominates electroweak symmetry breaking and the ETC coupling is small. A second set of solutions exist though in which technicolour plays a sub-dominant role to the ETC interactions which generate most of $f_\pi$ by being super-critical and generating masses that strongly break isospin in the technidoublet. These latter solutions are strongly ruled out by the $\delta \rho/T$ parameter so we do not explore them in much detail. The more normal solutions can be followed to larger top masses where the NJL interaction is strong. We find there is a maximum top mass (here the two branches of solutions merge) that is compatible with the electroweak scale which is a little above 500 GeV for a QCD-like, low $N_f$ model. For models with larger $N_f$ the enhancement of the techniquark condensate by walking allows a given top mass to be generated with a weaker ETC coupling and significantly larger $m_t$ can be achieved. These results confirm the ability to compute with both walking and strong NJL interactions present.

We then concentrate on models with $m_t=175$ GeV. We track the growing strength of the NJL coupling with rising ETC scale. Walking's enhancement of the condensate  allows solutions at lower ETC coupling for a given ETC scale. Phenomenologically the key question is whether these solutions are compatible with the tight $\delta \rho$ parameter constraint (it must be less than 0.4$\%$). There are two contributions to $\delta \rho$ \cite{Chivukula:1995dc,Appelquist:1996kp}. The first is a direct contribution in which a single ETC gauge boson is exchanged across a techniquark loop contributing to the W and Z masses. The contribution to $\delta \rho$ is expected to be
\begin{equation} \delta \rho =  {g^2 v^4 \over \Lambda^2}  \label{deltarho2loop} \end{equation}
where here $g$ also includes any group theory factors from the ETC model. This bound can be evaded by pushing the ETC mass scale up above 3 TeV or so although it is easier to avoid in walking (large $N_f$) models where the ETC interactions can be smaller. A second contribution is harder to avoid though \cite{Appelquist:1989ps}. The isospin breaking ETC interactions tend to generate mass splitting between the techni-U and techni-D quarks. This mass splitting gives $\delta \rho$ contributions. For a perturbative doublet, with $N_{TC}$ degeneracy, this mass splitting gives
\begin{equation} \delta \rho = 0.4\% N_{TC} \left( {\Delta m^2 \over (175 GeV)^2} \right) \label{deltarho3loop}\end{equation}
The holographic model allows us to plot the self energy function of the quarks against RG scale. We find typical mass splittings between 20 and a few 100 GeV. Interestingly extreme walking models generate the largest IR mass splitting. When the technicolour interactions are strong at the ETC scale the dynamics are much more sensitive to the high scale NJL isospin violation. Models with $N_f=3-8$ are compatible with both the $\delta \rho$ bounds as estimated so far for ETC scales out to 30TeV or above. One would hope that the holographic model would allow a non-perturbative estimate of $\delta \rho$ to move beyond (\ref{deltarho3loop}). This is a little subtle because holographically mixed flavour states are described by strings. For very small splittings the non-abelian DBI action \cite{Myers:1999ps,Erdmenger:2007vj} of a collection of branes would give a field theoretic computation for these states in which the background metric becomes some average over the two flavour embeddings. It is not clear this is valid for the large isospin breaking that is needed for the top  but we estimate $f_{\pi^\pm}$ in this fashion. The resulting computation shares much with Pagel Stokar type estimates \cite{Pagels:1979hd} depending not just on the value of the self energy but also its derivatives. Here that enlarges the $\delta \rho$ estimates substantially (by as much as an order of magnitude) and the maximum ETC scale compatible with the constraints lies between 5 and 15TeV depending on $N_f$. A judicous choice of a low ETC scale ($\sim$5 TeV), some walking ($N_f=8$), and strong ETC does appear compatible.   The tension with $\delta \rho$ has, of course, been previously observed (although we hope the holographic model provides a more robust framework for the observation) and was the motivation for top condensation assisted technicolour \cite{Miransky:1988xi,Miransky:1989ds,Nambu:1989jt,Bardeen:1989ds}. Here a separate NJL interaction is introduced for the top quark to generate its mass independently of the electroweak breaking technicolour sector which removes the isospin breaking from the technicolour sector. We briefly show this mechanism at work in the holographic model where the ETC interaction can be switched off as the top condensation coupling grows whilst still achieving a fixed $m_t$. 

Finally for completion we consider a one family ETC model with an SU(3) technicolour group, $N_f=8$ (there are now 4 electroweak doublets so the strain on S would be high!).  We compute the ETC coupling as a function of  ETC scale. The model faces worse constraints on the mass splitting in the techni-doublet since there are three colours of techni-U quarks. The holographic description does though allow the model to evade these constraints for ETC scales between 3 and 7 TeV.

\section{Dynamic AdS/QCD}

In this section we review the Dynamic AdS/QCD model \cite{Alho:2013dka} which we will use to describe the technicolour (and QCD) dynamics. The model is based on holographic ``top-down'' D7 probe models of chiral symmetry breaking \cite{Karch:2002sh,d3d7,Babington:2003vm,Filev:2007gb}. The models are surprisingly simple with a single field (the brane embedding) describing the quark condensate. The dynamics of the gauge theory manifests in the Dirac Born Infeld (DBI) action of the probe brane  as  scale (radially) dependent mass squared for the field.  Chiral symmetry breaking occurs if there is a violation the BF bound \cite{Jarvinen:2011qe,Kutasov:2011fr,Alvares:2012kr}. This occurs when the anomalous dimension of the quark bilinear grows to one. It is natural model building to replace the running of the mass squared with a phenomenological guess to realize the phenomenology of a wider range of theories which is our approach here (at the level of the DBI action this could be done by picking a form for the background dilaton field for example). 

The essential dynamics of the model is encoded in a field $X$ of mass dimension one. The modulus
of this field describes
the quark mass and condensate. Fluctuations in $|X|$ around its vacuum configurations will describe the higgs-like $\sigma$ meson. The $\pi$ fields are the phase of $X$
\begin{equation} X = L(\rho)  ~ e^{2 i \pi^a T^a} .
\end{equation}
Here $\rho$ is the holographic coordinate ($\rho=0$ is the IR, $\rho \rightarrow \infty$ the UV), 
and $|X|=L$ enters into the effective radial coordinate in the space, i.e. $r^2 = \rho^2 + |X|^2$. This allows the quark condensate to generate a soft IR wall: when $L$ is nonzero the theory will exclude the deep IR at $r=0$. This implementation is taken directly from the D3/probe-D7 model  where $L$ is the embedding of the D7 brane in the AdS spacetime. Fluctuations on the brane then see the pulled back metric on the D7 world volume.

We work with the five dimensional metric 
\begin{equation} 
ds^2 =  { d \rho^2 \over (\rho^2 + |X|^2)} +  (\rho^2 + |X|^2) dx^2, 
\end{equation}
which will be used for contractions of the space-time indices.The five dimensional action of our effective holographic theory is
\bea
S & = & \int d^4x~ d \rho\, {\rm{Tr}}\, \rho^3 
\left[  {1 \over \rho^2 + |X|^2} |D X|^2 \right. \nonumber \\ 
&& \left.+  {\Delta m^2 \over \rho^2} |X|^2   + {1 \over 2 \kappa^2} (F_V^2 + F_A^2) \right], 
\label{daq}
\eea
where $F_V$ and $F_A$ are vector fields that will describe the vector ($V$) and axial ($A$) mesons.
Note that we have not written the $\sqrt{-g}$ factor in the metric as $r^3$ but just $\rho^3$. Again, this is driven by the D7 probe action in which this factor is $\rho^3$;  maintaining this form is crucial to correctly implementing the soft wall behaviour. 
Finally $\kappa$ is a constant that will determine the $V-A$ mass splitting and enter into the $f_\pi$ computation; we will fix its value and $N_f$ dependence in our model below. The model presented is phenomenological in nature and we have included the bare minimum of content to reproduce the broad physics we expect. Thus, for example, we include a mass term for $X$ so that we may encode the running of the anomalous dimension of the quark bilinear but we neglect higher order terms in $X$.

The vacuum structure of the theory is found by setting all fields except $|X|=L$ to zero. We further assume that $L$ will have no dependence on the $x$ coordinates. The action for $L$  is given by
\begin{equation} \label{act} S  =  \int d^4x~ d \rho ~  \rho^3 \left[   (\partial_\rho  L)^2 +  \Delta m^2 {L^2  \over \rho^2 }   \right].
\end{equation}
Now if we re-write $L =  \rho \phi $ and integrate the first term by parts we arrive at
\begin{equation}  S  =  \int d^4x~ d \rho ~ ( \rho^5  (\partial_\rho  \phi)^2 +  \rho^3 (-3+ \Delta m^2) \phi^2  )\,,
\end{equation}
which is the form for a canonical scalar in AdS$_5$. The usual AdS relation between the scalar mass squared and the dimension of the field theory operator applies ($m^2 = \Delta (\Delta-4)$).   If $\Delta m^2 =0 $ then the scalar describes a dimension 3 operator and dimension 1 source as is required for it to represent $\bar{q} q$ and the quark mass $m$. In the UV the solution for the $\phi$ equation of motion is $\phi = m/\rho + \bar{q}q/\rho^3$.

The Euler-Lagrange equation for the determination of $L$, in the case of a constant $\Delta m^2$, is 
\begin{equation} \label{embedeqn}
\partial_\rho[ \rho^3 \partial_\rho L]  - \rho \Delta m^2 L  = 0\,. \end{equation}
We can now ansatz an $r$ dependent $\Delta m^2$ to describe the running of the dimension of $\bar{q}q$ (we do this at the level of the equation of motion). If the mass squared of the scalar violates the BF bound of -4 ($\Delta m^2=-1$) then we expect the scalar field $L$ to become unstable and settle to some non-zero value. 
To enact a realization of various gauge theories we will use the perturbative running from SU($N_c$) gauge theories with $N_f$ flavours since the two loop results display a conformal window - this is where we include the dynamics of a particular gauge theory.  

The two loop running of the gauge coupling in QCD is given by
\begin{equation} 
\mu { d \alpha \over d \mu} = - b_0 \alpha^2 - b_1 \alpha^3,
\end{equation}
where
\begin{equation} b_0 = {1 \over 6 \pi} (11 N_c - 2N_F), \end{equation}
and
\begin{equation} b_1 = {1 \over 24 \pi^2} \left(34 N_c^2 - 10 N_c N_f - 3 {N_c^2 -1 \over N_c} N_F \right) .\end{equation}
Asymptotic freedom is present provided $N_f < 11/2 N_c$. There is an IR fixed point with value
\begin{equation} \alpha_* = -b_0/b_1\,, \end{equation}
which rises to infinity at $N_f \sim 2.6 N_c$. 

The one loop result for the anomalous dimension is
\begin{equation} \gamma = {3 C_2 \over 2\pi}\alpha= {3 (N_c^2-1) \over 4 N_c \pi} \alpha\,.  \end{equation}
So, using the fixed point value $\alpha_*$, the condition $\gamma=1$ occurs at $N_f^c \sim 4N_c$ (this is the edge of the conformal window in the model).

We will identify the renormalization group (RG) scale $\mu$ with the AdS radial parameter $r = \sqrt{\rho^2+L^2}$ in our model. Note it is important that $L$ enters here. If it did not and the scalar mass was only a function of $\rho$ then were the mass to violate the BF bound at some $\rho$ it would leave the theory unstable however large $L$ grew. Including $L$ means that the creation of a non-zero but finite $L$ can remove the BF bound violation leading to a stable solution. Again this has a natural origin in the D3/D7 system.

Working perturbatively from the AdS result $m^2 = \Delta(\Delta-4)$ we have
\begin{equation} \label{dmsq3} \Delta m^2 = - 2 \gamma = -{3 (N_c^2-1) \over 2 N_c \pi} \alpha\, .\end{equation}
This will then fix the $r$ dependence of the scalar mass through $\Delta m^2$ as a function of $N_c$ and 
$N_f$. 

To find numerical solutions we need an IR boundary condition. In top down models $L'(0)=0$ is the condition for a regular solution.  Since we do not wish to describe IR physics below the quark mass (where the quark contribution to the running coupling will decouple) we use a very similar on-shell condition - we shoot from points $L(\rho=L_0) = L_0$ with $L'(L_0)=0$. 


The spectrum of the theory is determined by looking at linearized fluctuations of the fields about the vacuum. 
The normalizations of the fluctuations are determined by matching to the gauge theory in the UV of the theory. External currents are associated with the non-normalizable modes of the fields in AdS. In the UV we expect  $|X| \sim 0$ and we can solve
the equations of motion for the scalar, $L= K_S(\rho) e^{-i q.x}$, vector $V^\mu= \epsilon^\mu K_V(\rho) e^{-i q.x}$, and  axial $A^\mu= \epsilon^\mu K_A(\rho) e^{-i q.x}$ fields. Each satisfies the same equation
\begin{equation}  \label{thing}
\partial_\rho [ \rho^2 \partial_\rho K] - {q^2 \over \rho} K= 0\,. \end{equation}
The UV solution  is
\begin{equation} \label{Ks}
K_i = N_i \left( 1 + {q^2 \over 4 \rho^2} \ln (q^2/ \rho^2) \right),\quad (i=S,V,A),
\end{equation}
where $N_i$ are normalization constants that are not fixed by the linearized equation of motion.
Substituting these solutions back into the action gives the scalar correlator $\Pi_{SS}$, the vector correlator $\Pi_{VV}$ and axial vector correlator $\Pi_{AA}$. Performing the usual matching to the UV gauge theory requires us to set
\begin{equation} N_S^2 = {N_c N_f \over 24 \pi^2 }, \hspace{0.5cm} N_V^2 = N_A^2 = {\kappa^2 N_c N_f \over 24 \pi^2 }.
\end{equation}

As an example, the axial meson spectrum is determined from the equation of motion for the spatial pieces of the axial-vector gauge field. In the $A_z=0$ gauge we write $A_\mu= A_{\mu \perp} + \partial_\mu \phi$. 
The appropriate equation with $A_{\mu \perp} = \epsilon^\mu A(\rho) e^{-i q.x}$ with $q^2=-M^2$ is
\begin{equation}  \label{aa}  \partial_\rho \left[ \rho^3 \partial_\rho A \right]   - \kappa^2 {L_0^2 \rho^3 \over  (L_0^2 + \rho^2)^2} A + {\rho^3 M^2 \over (L_0^2 + \rho^2)^2} A = 0\,. \end{equation}
We again impose $A'(0)=0$ in the IR and require in the UV that $A\sim c/\rho^2$. To fix $c$ we normalize the wave functions such that the vector meson kinetic term is canonical
\begin{equation}  \int d \rho {\rho^3  \over \kappa^2 (\rho^2 + L_0^2)^2} A^2 = 1\,. \end{equation}

We fix $\kappa$ following the discussion in \cite{Clemens:2017udk}. 
\begin{equation} \kappa^2 = 7.6 (N_f- N_f^c). \label{kchoice}\end{equation}
The numerical factor gives a sensible fit to QCD with $N_c=3, N_f=2$ and the $N_f$ dependence is assumed to restore $\rho$ $a$ degeneracy at the edge of the conformal window. This latter condition is not clear cut but helps reduce the electroweak S parameter in walking technicolour. The choice is not crucial for our analysis below since we do not tune extremely close to the edge of the conformal window.

The pion decay constant can be extracted from the expectation that $\Pi_{AA} = f_\pi^2$. From the $f_A$ kinetic term with two external (non-normalizable) axial currents at $Q^2=0$ we obtain
\begin{equation} f_\pi^2 = \int d \rho {1 \over \kappa^2}  \partial_\rho \left[  \rho^3 \partial_\rho K_A(q^2=0)\right] K_A(q^2=0)\,.
\end{equation} 

For other states and decay constants the procedure is given in detail in \cite{Alho:2013dka}.

\section{NJL Operators}

We will wish to introduce four fermion operators into the technicolour models we will study to feed the techniquark condensate down to give the top quark a mass. At least in some cases these operators will be near or super critical in the sense of the Nambu Jona-Lasinio model \cite{Nambu:1961tp}. Here we will very briefly review the NJL model and show how to enact it in our holographic setting \cite{Evans:2016yas}.

Consider a free fermion with a four fermion interaction $g^2/\Lambda^2 \bar{q}_L q_R \bar{q}_R q_L$. In the standard NJL approximation there are two contributions to the effective potential. First there is the one loop Coleman Weinberg potential for the free quarks 
\begin{equation} V_{\rm eff} = - \int^\Lambda_0 {d^4k \over (2 \pi)^4} Tr \log (k^2 +m^2) \label{ColemanW}\end{equation}
This falls with growing $m$ and is unbounded, although normally one treats $m$ as a fixed parameter so one would not seek to minimize this term. When we add the four fermion term we allow $m$ to become dynamically determined but there is the extra term from the four fermion interaction evaluated on $m= (g^2/\Lambda^2) \langle \bar{q} q \rangle$
\begin{equation} \Delta V_{\rm eff} = {\Lambda^2 m^2 \over g^2} \label{NJLextra} \end{equation}
This makes the effective potential bounded and ensures a minimum. For small $g$ the extra term is large and the minimum is at $m=0$. When $g$ rises above $2 \pi$ the minimum lies away from $m=0$ and is given by the ``gap equation'' condition
\begin{equation} 
1 = {g^2 \over 4 \pi^2} \left( 1 - {m^2 \over \Lambda^2} \log \left[{\Lambda^2 + m^2 \over m^2 }\right] \right)
\end{equation}
The phase transition is second order. 

Next we will understand how to include the same NJL operator in a holographic model using Witten's multi-trace operator prescription \cite{Witten:2001ua}. Consider the Dynamic AdS/QCD model with no running ($\Delta m^2 = 0$) 
\begin{equation} S  =  \int d^4x~ d \rho ~  \rho^3   (\partial_\rho  L)^2. \label{norun}
\end{equation}
 Varying the action gives
\begin{equation}  \delta S = 0 = \int d \rho \left(\partial _\rho{\partial {\cal L }\over \partial L'}  - {\partial {\cal L }
\over \partial L} \right)  \delta L  + \left. {\partial {\cal L }\over \partial L'} \delta L \right|_{{UV, IR}} \,.  \end{equation}
Since the action only depends on $L'$ there is a conserved quantity $-2c$ from which we learn
\begin{equation} L' = { - 2 c \over \rho^3} \label{conserved} \,, \end{equation}
and hence  
\begin{equation}  L = m + {c \over \rho^2} \,. \end{equation}
The standard holographic interpretation is that $m$ represents a source, here the quark mass, and $c$ represents the operator $\bar{q} q$ condensate.

Normally one fixes $m$ in the UV as a parameter of the theory so $\delta L|_{UV}=0$ and then we require $ {\partial {\cal L} \over \partial L'}|_{IR}  =
\rho^3 L' =0 $ which is satisfied when $L'=0$. The equation of motion and both UV and IR boundary conditions vanish. We have arrived at the solution $L=m$.  

Witten's prescription for including the NJL operator is simply to require at the UV scale that $m= {g^2 \over \Lambda^2} c$. We can achieve this by adding a UV boundary action term
\begin{equation} \Delta S _{UV}=   {L^2 \Lambda^2 \over g^2  } \label{bound} \,.  \end{equation}
Now at the UV boundary we no longer require after variation of $L$ $\delta L=0$ but allow $L$ to change and instead impose 
\begin{equation} 0 = {\partial {\cal L} \over \partial L'}  + {2 L \Lambda_{UV}^2 \over g^2 }  \,,   \end{equation}
which gives the required $c,m$ relation at leading order for large $\Lambda_{UV}$ where $L \simeq m$.
The prescription maintains the IR boundary condition $L'=0$.  Note that the term added to the effective potential (\ref{bound}) with $L \simeq m$ is exactly that in (\ref{NJLextra}).

Now in the model of (\ref{norun}) the solution $L=m$ still solves the equation of motion and it still satisfies the IR boundary condition when $c=0$. The only solution that then satisfies the UV boundary condition is $m=c=0$. The fact that however large $g$ is $m=0$ is the only solution is the puzzle that \cite{Evans:2016yas} resolved. It is clear from the effective potential: the action (\ref{norun})  evaluated on $L=m$ vanishes for all $m$. Interpreting this as the effective potential and adding (\ref{NJLextra}) clearly leads to the minimum $m=0$. The point is that the action (\ref{norun}) has failed to reproduce (\ref{ColemanW}). The reason is that the simple model has been taken from an N=2 supersymmetric construction in which the vacuum energy vanishes for all theories whatever the quark supermultiplet mass is. In \cite{Evans:2016yas} it was shown that supersymmetry breaking in the N=2 model leads to a non-trivial potential from the bulk and standard NJL behaviour returns. In \cite{Clemens:2017udk} we set the scene for the alnalysis here by breaking supersymmetry by the running of $\Delta m^2$ in Dynamic AdS/QCD, representing the gauge dynamics, and showed that in the presence of an NJL term the NJL transition is smoothed from first to second order. The UV NJL term enhances the IR symmetry breaking of the gauge theory enlarging the mass gap.  Formally the absence of (\ref{ColemanW}) might look serious in a weakly coupled theory but at strong coupling the effective potential will be dominated by loops with gluon exchange (``cracked egg'' diagrams) and our models will include these. In the next section we look at the simplest example where the NJL model, rather than the gauge dynamics, is responsible for the bulk of chiral symmetry breaking.

\section{A Hologram of Top Condensation}

The simplest model of NJL operators within Dynamic AdS/QCD is top condensation. We consider the case with the quark anomalous dimension running with $N_c=3$ and $N_f=6$ massless quarks to represent the six standard model quarks and their QCD interactions. This running breaks the conformal symmetry of the model and introduces a bulk contribution to the effective potential in analogy to (\ref{ColemanW}). We set $\alpha_s(e GeV)=0.39$ so the BF bound is violated at 1 GeV setting the scale $\Lambda_{QCD}$. Without an NJL operator  the strong force becomes strong at the few hundred MeV scale where it breaks chiral symmetry and generates an IR quark mass for all six quarks of $\sim 350$MeV. 

We will then include the four fermion interaction
\begin{equation} {g^2 \over \Lambda^2} \left(\bar{\psi}_L t_R \bar{t}_R \psi_L ~~ + ~~ h.c. \right) \end{equation}
Note this is for one flavour, the top, only; $\psi_L$ is the SU(2)$_L$ top-bottom multiplet but only the top quark mass is influenced since only $t_R$ enters. 

\begin{figure}[]
\centering
\includegraphics[width=8cm]{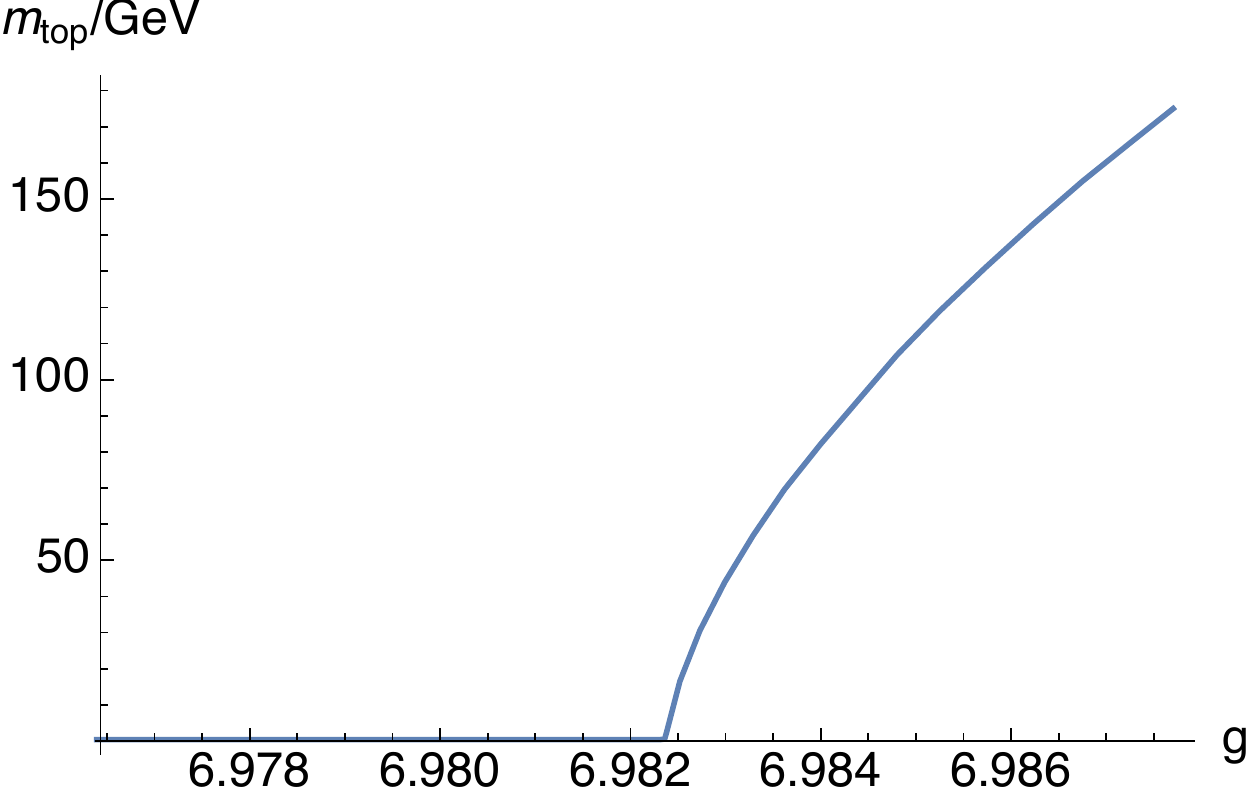}
\caption{The top condensation model with  $\Lambda=10TeV$: the IR top mass against NJL coupling showing critical value of the coupling. Note below the critical value the mass rises from 365 MeV at $g=0$ from the underlying QCD dynamics. }
\label{topmass}
\end{figure}

To impose the presence of the NJL operator in the holographic model we require that the embedding function for the top quark at the cut off $\Lambda$ takes the form
\begin{equation} L_t = m_t + {c_t \over \rho^2}, \hspace{1cm}  m_t = {g^2 c_t \over \Lambda^2}  \label{topcon}\end{equation}
To numerically extract $m_t$ and $c_t$ we perform a numerical fit of this form to $L(\rho)$ in a small range in $\rho$ just below $\Lambda$.

The 5 remaining quarks play only a spectator role contributing to the form of the running of $\gamma$. For these we require $L \rightarrow 0$ at the cut off so they are massless. They also make a negligible contribution to the electroweak $f_\pi$ of order 100 MeV. 

We proceed by picking the IR boundary value of $L_t$ for the top, which we interpret as the IR value of the top mass, $m_t^{\rm phys}$. We then numerically evolve by shooting to the UV boundary $\Lambda$. There we can read off the UV values of $m_t$ and $c_t$ and impose the NJL condition (\ref{topcon}) to extract $g$.  In Fig \ref{topmass} we show the resulting plot for 
$\Lambda=10$TeV. It shows the classic NJL behaviour of the presence of a critical value of the coupling at which a second order transition occurs (in fact because of the underlying QCD dynamics the IR mass does not fully switch off below the critical coupling but it does fall to just 350MeV). To achieve $m_t^{\rm phys} \ll \Lambda$ requires one to live fine tuned to the critical coupling as one would expect. It's interesting that here, because we choose the IR top mass, numerically there is no difficult tuning to be done - it emerges once one computes $g$ for those solutions.

\begin{figure}[]
\centering
\includegraphics[width=8cm]{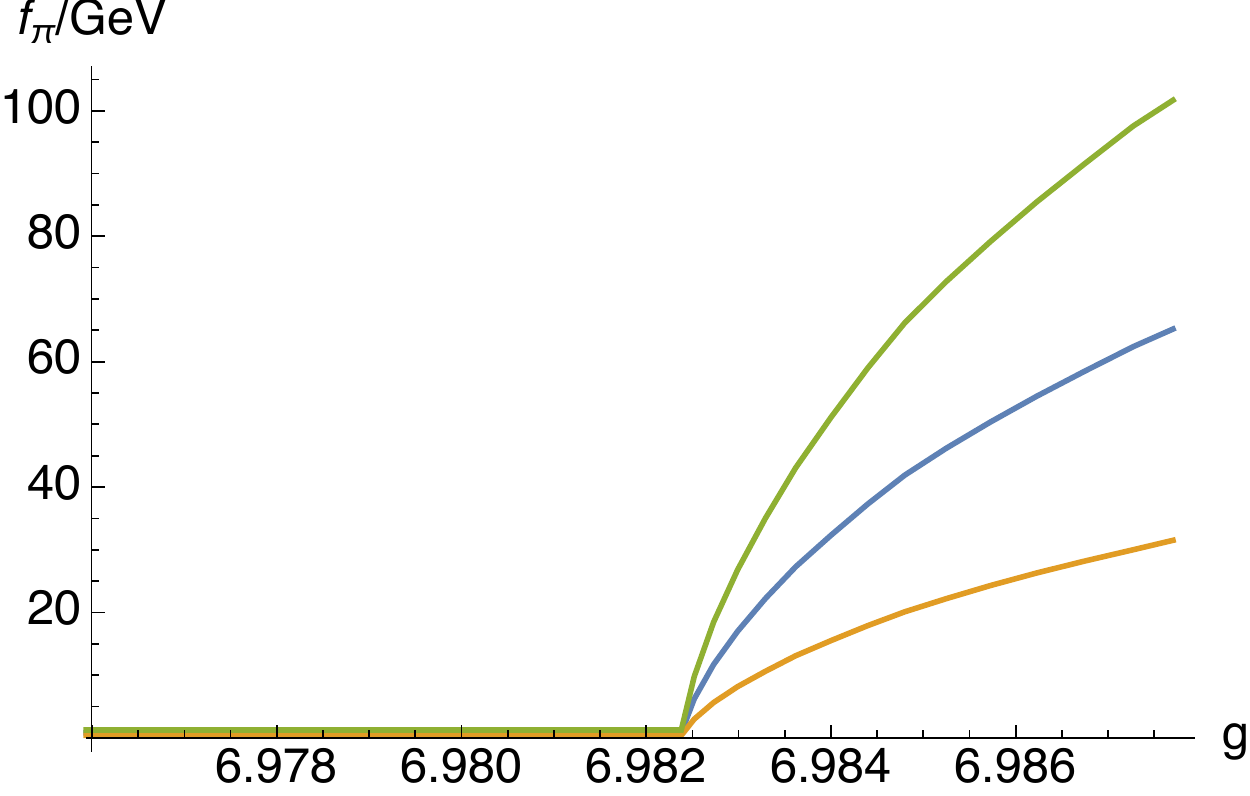}
\caption{The top condensation model: $\Lambda=10TeV$: top contribution to $f_\pi$ against NJL coupling showing critical value of the coupling. Here for $\kappa=1,5,15$ from bottom to top. }
\label{fpitop}
\end{figure}

\begin{figure}[]
\centering
\includegraphics[width=8cm]{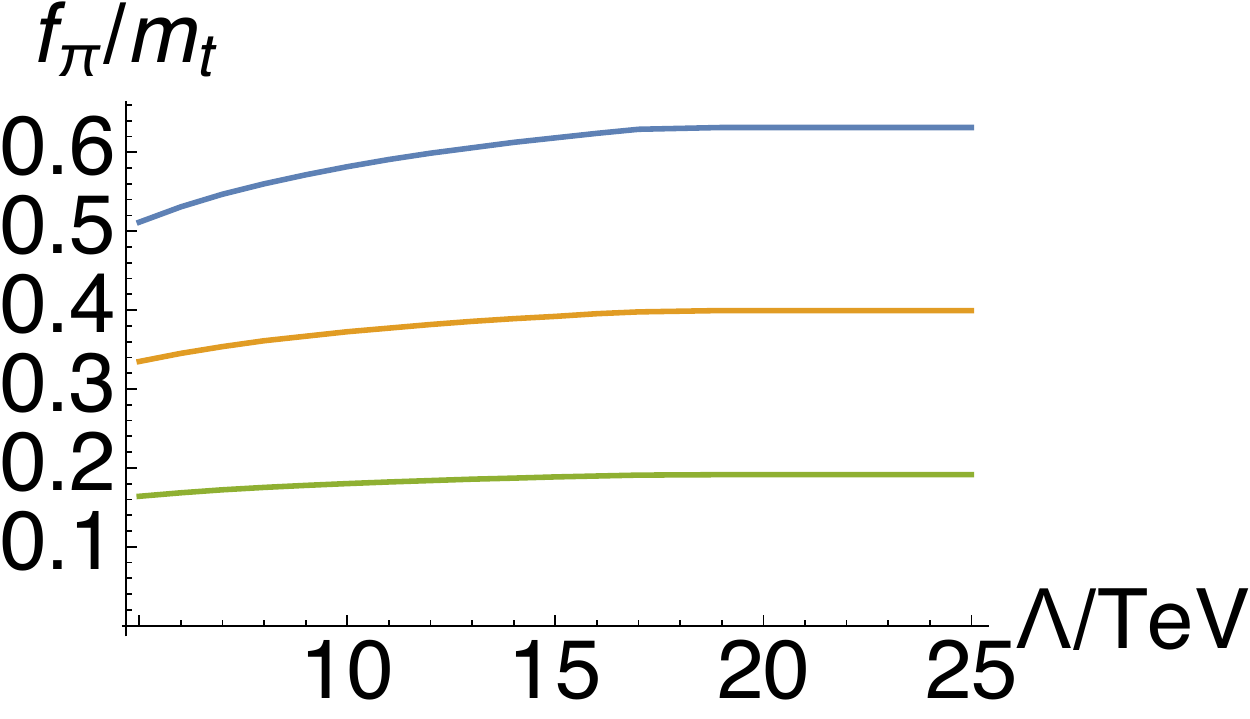}
\caption{The top condensation model tuned to $m_t^{\rm phys}=175$ GeV: $f_\pi$/$m_{\rm top}$ against $\Lambda$ for $\kappa=1,5,15$ from bottom to top. }
\label{fpitoplam}
\end{figure}

In Fig \ref{fpitop} we show the resulting computation of $f_\pi$ in this model. It again shows critical NJL behaviour. The precise value depends on the choice of the parameter $\kappa$. One might usually fix $\kappa$ from the $\rho-a$ mass splitting in QCD but this is a low energy estimate of $\kappa$ which could change by scales as enormous as 10 TeV. The usual expectation in top condensation models is that the observed top mass (which corresponds to the largest values of $g$ shown in Figs \ref{topmass} \& \ref{fpitop}) is insufficent to generate the electroweak $f_\pi$ and this is bourne out here by choices of $\kappa$ of order one as shown. Henceforth in describing the top sector we will take $\kappa=1$. In Fig  \ref{fpitoplam} we display the dependence of $f_\pi$  on the cut off scale - here at each value of $\Lambda$ we have arranged $g$ to generate the physical top mass. The value asymptotes to fixed values with higher $\Lambda$. For $\kappa \simeq1$ $f_\pi$ can not achieve the electroweak scale, the usual failure of top condensation.

We can also compute the mass of the scalar bound state of the top quark, $\sigma$, and we find the value of its mass is very stable with $\Lambda$ at $ \simeq 590$GeV. This is the state which in a pure top condensation model would correspond to the higgs boson.

\section{A Hologram of One Doublet ETC}

We will next describe a ``classic'' dynamical model of the top mass - technicolour plus extended technicolour interactions to generate the top mass. 

Consider a model with an SU($N_{TC}$) gauge group under which a single electroweak doublet of techniquarks $(U,D)$ transform in the fundamental representation. In addition there may be extra electroweak singlet techniquarks that allow us to dial $N_f^{\rm sing}$ and hence change the running. This sector is simply described by Dynamic AdS/QCD with the running fixed by $N_{TC}$ and $N_f = 2 + N_f^{\rm sing}$. The only remaining freedom (given the choice of $\kappa$ in (\ref{kchoice})) is the value of $\alpha_{TC}$ at some scale (we have found it numerical useful to set this paramter at the scale  $e^2$ TeV) which one dials to generate the correct $f_\pi$ for electroweak symmetry breaking. A naive model such as this of the technicolour sector preserves custodial isospin - the ETC sector will break that.

A simple ETC model places the top quark ($t_R$ and $\psi_L=(t,b)_L$) and the techni-quarks ($U_R$ and $\Psi_L=(U,D)_R$) in the fundamental representation of an SU(6) ETC gauge group that is broken at some scale (by dynamics we don't specify) to SU(3)$_{TC} \otimes$ SU(3)$_{QCD}$ generating a mass, $\Lambda$, for the ETC gauge bosons associated with the broken generators. The ETC boson exchange associated with broken step up and down operators of the SU(6) ETC group form the four fermion operators
\begin{equation} {g^2 \over 2 \Lambda_{ETC}^2} \bar{\Psi}_L^\alpha U_R^\alpha \bar{t}_R^i \psi_L^i \end{equation}
here $\alpha$ is a technicolour index and $i$ is a QCD colour index each of which are summed over. There is also a broken diagonal generator\\ \\
\centerline{   $T_{\rm diag} = {1 \over \sqrt{12} } {\rm diag}(1,1,1,-1,-1,-1)$} \\ \\
 which gives us
\begin{equation} {g^2 \over 12 \Lambda^2} \bar{\Psi}_L^\alpha U_R^\alpha \bar{U}_R^\beta \Psi_L^\beta  +
 {g^2 \over 12 \Lambda^2} \bar{\psi}_L^i t_R^i \bar{t}_R^j \psi_L^j  \end{equation}

Holographically we will describe the QCD quark sector including the top as in the top condensation model: we take a second Dynamic AdS/QCD sector with $N_c=3$ to represent QCD and $N_f=6$ to represent the six quarks. We set $\alpha$ so that the BF bound is violated at the 1 GeV scale to represent QCD becoming strongly coupled. In this model we solve numerically for the embedding function $L(\rho)$ for the top quark subject to $L_t^{IR}=m_t^{\rm phys}$. This function is now fixed and from it we can read off the UV embedding parameters $m_t$ and $c_t$ at any scale $\Lambda$ by fitting to the form $L \sim (m_t + c_t/\rho^2)|_{\Lambda}$. The remaining quark masses are so small that we leave them as massless spectators at the electroweak scale. 

In the Dynamic AdS/QCD description of the technicolour sector we now split the embedding functions for the U and D techniquarks. The D quark's embedding function, 
$L_D(\rho)$, must fall to zero at the UV cut off scale - we will find this unique function for each choice of $\alpha_{TC}(e^2 TeV)$ and $\Lambda$. The U techniquark embedding function $L_U$ though will be allowed to  have non-zero $m_{U}$ at the UV scale and we will read off $m_U$ and $c_U$ in the same fashion as for the top. For each choice of the IR value of $L_U$, which leads to a UV pair $(m_{U},c_{U})$, we must also pick $\alpha_{TC}(e^2 TeV)$ so that the sum of the U,D and top contributions to $f_\pi$ match the electroweak scale ($f_\pi$=246 GeV). Alternatively one can chose a value of $\alpha_{TC}(e^2 TeV)$ and allow $L_U^{IR}$ to vary to match $f_\pi$.

\begin{figure}[]
\centering
\includegraphics[width=8cm]{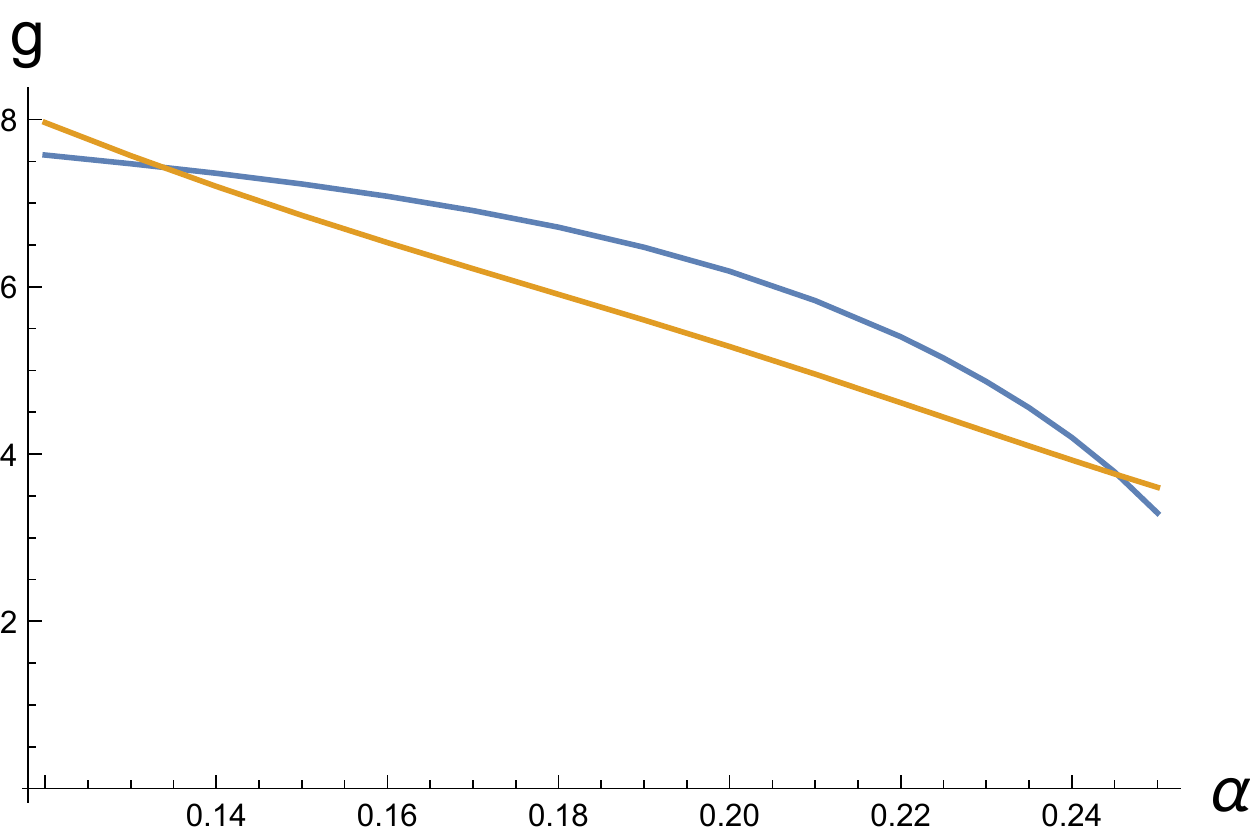}
\caption{One doublet model ($N_{TC}=3, N_f=2$) with $\Lambda=5TeV$. We use an embedding for the top quark with $L_t^{IR}=175$ GeV. We vary $\alpha_{TC}(e^2$ TeV) and then determine $L_D(\rho)$ that vanishes at the cut off, and the value of $L_U^{IR}$ that ensures the correct EW $f_\pi$. We then plot the value of $g$ from each of (\ref{g1}) and (\ref{g2}). The crossing points mark a self consistent solution and determines $g$. The left point is an NJL dominated solution the right hand one TC dominated.  }
\label{gtune}
\end{figure}

The job now is to find the choice of  $\alpha_{TC}(e^2 TeV)$ and $m_{U}$ at the UV scale that is consistent with the desired top mass  given the ETC interactions we have chosen. Holographically the multi-trace prescription for our NJL operators are
\begin{equation}  m_{U} = {g^2  \over 12 \Lambda^2} c_{U} + {g^2 \over 2 \Lambda^2} c_{t} \label{g1}\end{equation}
and 
\begin{equation}  m_{t} = {g^2 \over 12 \Lambda^2} c_{t} + {g^2 \over 2 \Lambda^2} c_{U} \label{g2} \end{equation}

\begin{figure}[]
\centering
\includegraphics[width=8cm]{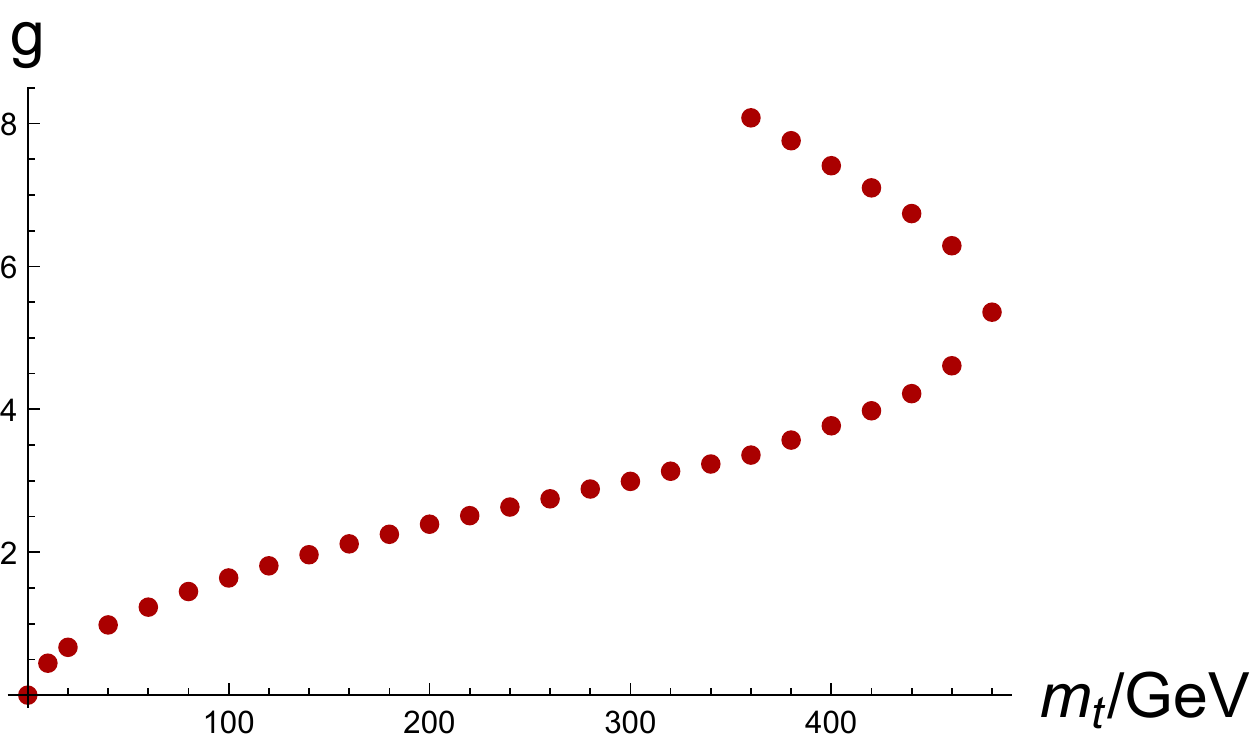}
\caption{Plots of $g$ vs $m_t^{\rm phys}$ for consistent solutions in the one doublet model with $N_c=3$ and $N_f=2$ at $\Lambda=$5 TeV showing both TC and NJL dominated branches.     }
\label{gvsmtnf2}
\end{figure}

Thus at each choice of $\Lambda$ we must plot the value of $g$ extracted from each of these equations as we vary the $L_U^{IR}/\alpha_{TC}(e^2 TeV)$ pair, each time getting different $(m_U,c_U)$ pairs. We seek the point where both equations return the same value of $g$ and are self consistent. An example of this fit is shown in Fig \ref{gtune}. Note that generically there are two solutions. The left hand cross point at higher $g$ is an ``NJL dominated'' model of electroweak symmetry breaking - the technicolour interaction is rather weak and the technidown quark plays almost no role in generating the electroweak $f_\pi$. The top and techni-up quark are both heavy and contribute dominantly to the electroweak scale. These solutions, whilst interesting, are at odds with experiment. They have very large isospin breaking between the U and D techniquarks which is certainly ruled out experimentally. The right hand solution at lower ETC coupling is a more technicolour dominated model. The techniquarks provide most of the electroweak scale and are, at least somewhat, degenerate. We will  concentrate on these latter solutions below. Note that as the cut off is increased or the desired top mass raised the two curves in Fig \ref{gtune} pass through each other - the two solutions move together and will eventually coalesce into a single solution before at higher $m_t$ or $\Lambda$ there is no physical solution. The critical solution is where both strong ETC and TC are working together hardest to generate the largest possible top mass whilst still maintaining the physical weak scale.   

In Fig \ref{gvsmtnf2} we show an example of the evolution of the two solutions with varying $m_t^{\rm phys}$. Here the model has $N_{TC}=3$ and $N_f^{sing}=0$ (thus a total $N_f=2$ model) and we solve for $g$ to generate different values of the top mass with an ETC scale of 5 TeV. We see that at generic $m_t^{\rm phys}$ there are two branches - the lower weakly coupled ETC branch merges to $g=0$ at $m_t^{\rm phys}=0$ and that is the standard weakly coupled ETC behaviour. For higher $m_t^{\rm phys}$ there are two solutions with one having a larger ETC coupling - these solutions are where the D's contribution to $f_\pi$  is much smaller than the U's. At $m_t^{\rm phys} \simeq 500$ GeV the two branches merge and this is the maximum achievable top mass in the model with these parameters (higher $m_t^{\rm phys}$ could be achieved if $f_\pi$ was raised above the physical value). Henceforth we will neglect the upper branch since it is phenomenologically unacceptable due to the huge isospin breaking in the techniquark sector. Note here the experimental top mass is achievable.

We are now ready to explore how $g$ must be chosen to generate the observed top mass for any given choice of $\Lambda_{ETC}$ and $N_f^{\rm sing}$.
Two mechanisms have been proposed for how to obtain the 175 GeV physical top mass with an ETC scale of a few TeV or above in this system. The first is to allow the ETC interactions to  become strong. The second is to enhance the techniquark condensate by walking dynamics. We can see both mechanisms at work here.

\begin{figure}[]
\centering
\includegraphics[width=8cm]{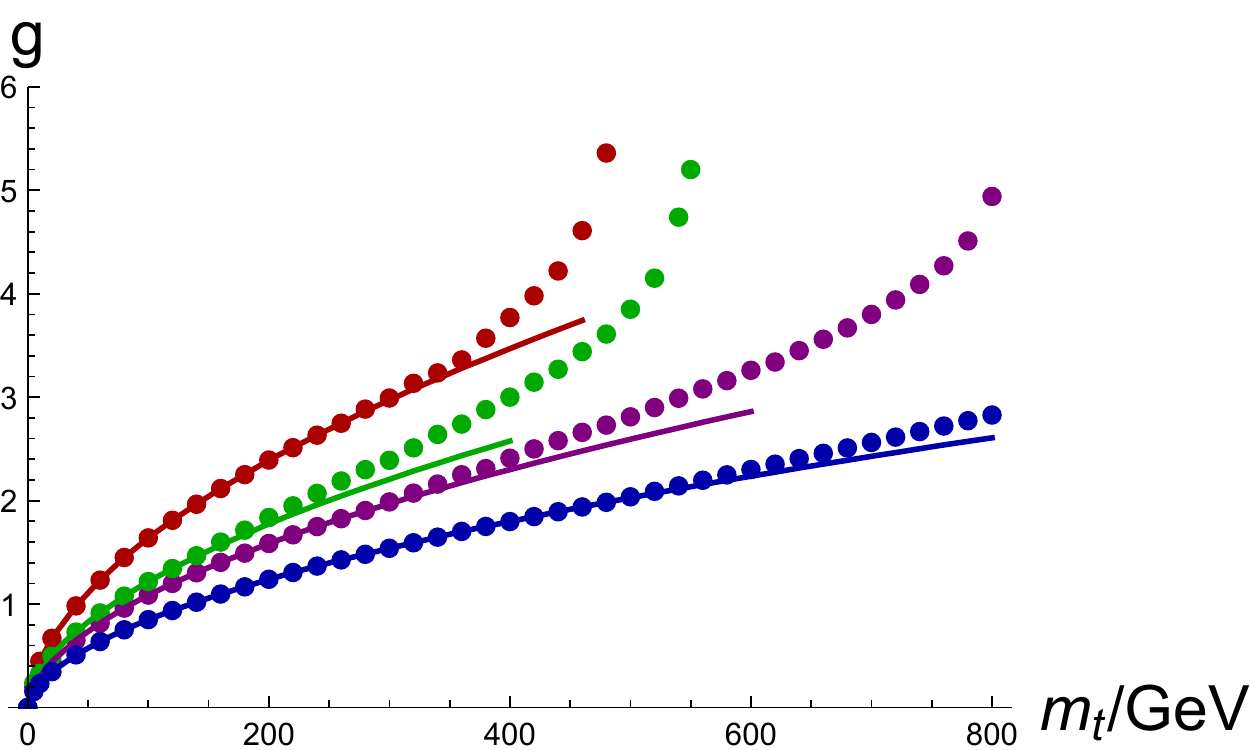}
\caption{Plots of $g$ vs $m_t^{\rm phys}$ in the one doublet model with $N_c=3$ and $N_f=2,4,8,11$ (from the top down). The points are data from the holographic model. The curves are the result of computing just using the simple ETC formula from just the last term in (\ref{g2}). For the first three cases the final point is the largest value of $m_t$ achievable.    }
\label{gvsmt}
\end{figure}

Let's again consider the model with $N_{TC}=3$ and $N_f^{\rm sing}=0$ which has a very running gauge coupling and so we expect to need to depend on strong ETC to generate the 175 GeV top quark mass. For smaller top mass values naively one would consider ETC to be weakly coupled and just use the last term in (\ref{g2}) and it is interesting to see how badly that approximation fares at large ETC coupling. To test this we can study the theory at  $\Lambda=5$ TeV -  again we fit for $g$ as a function of the top quark masses, $m_t^{\rm phys}$, in our model (in each case requiring the 246 GeV value of $f_\pi$). The results are repeated as the top curve/points in Fig \ref{gvsmt}. The solid line is the prediction of $g$ from just the final term in (\ref{g2}) - here we neglect the top mass in computing $f_\pi$ and determine the condensate in a fully isospin symmetric TC model. The points are the full data from our model. In fact they lie reasonably close except near the highest $m_t$ value - that highest value is a non-perturbative prediction of the model. Note at least a part of the reason that the full model requires a larger ETC coupling for high top mass is that the top is contributing significantly to $f_\pi$ which drives the TC scale and condensate down.

The walking argument \cite{Holdom:1981rm} says that if we tune $N_f$ to the critical number of flavours for chiral symmetry breaking  then the running will leave the theory with an anomalous dimension for the quark condensate close to 1 upto large scales approaching the ETC cut off. The dimension 3 condensate will then be given by the enlarged $\langle \bar{q} q \rangle \simeq 4 \pi v^2 \Lambda$. For $N_f=3$ the edge of the conformal window is just below 12 in the approximations we make. In Fig \ref{gvsmt} we repeat the above computation for $N_f=4,8$ and $11$ ($N_f^{\rm sing}=2,6,9$). These are, in order, the curves below the $N_f=2$ case in the figure. The enhancement of the condensate is apparent with the ETC coupling values falling by 2 or more as $N_f$ grows. If one allowed fractional $N_f$ values above 11 the condensate can be driven arbitrarily higher yet which reflects the ability to tune further if one introduced higher $N_c$ values.  The model does therefore incorporate the walking solution for generating the top mass too. 

We stress that independently of the phenomenology we will next discuss it is a success to be able to compute in a model that incorporates both strong NJL operators and the walking enhancement of the quark condensate.

\begin{figure}[]
\centering
\includegraphics[width=8cm]{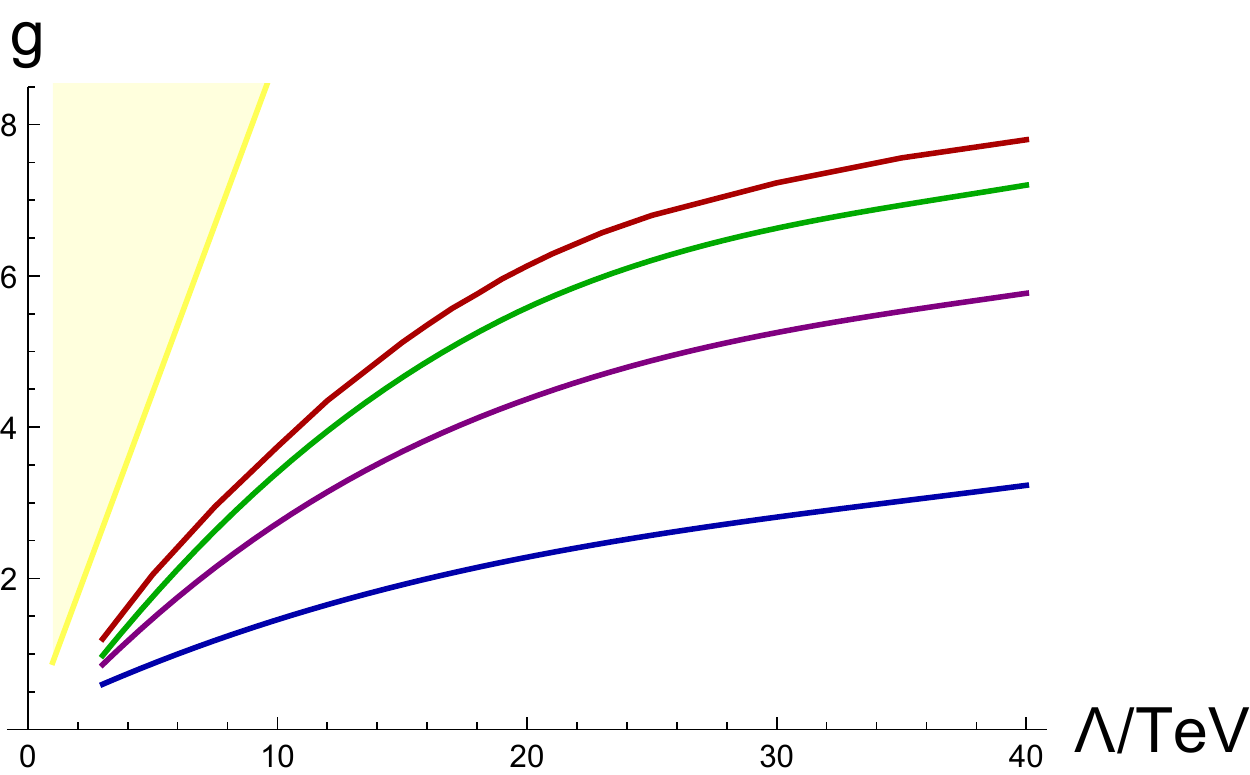}
\caption{ $g$ vs UV cut off $\Lambda$ for consistent solutions with the physical top mass on the TC dominated branch for $N_c=3$, $N_f=2,4,8,11$ from the top down. The shaded region is excluded by the two loop $\delta \rho$ contribution.  }
\label{gvsLam}
\end{figure}

Next we can study the ability of the theories to generate the experimental top mass at different ETC mass scales. Now fixing  $m_t^{\rm phys}=175$ GeV we search for the ETC coupling $g$ in each of the theories at different cut offs. We display the results in Fig \ref{gvsLam} - the curves are for $N_f=2,4,8,11$ coming down the plot. The plot again nicely illustrates the two mechanism at work here. The curves bend down to the right from straight because the strong ETC dynamics is enhancing the top mass. As $N_f$ increases the coupling needed to generate the phyiscal top mass falls because walking is enhancing the condensate.

Are these solutions phenomenologically acceptable though? The worry as we stressed in the introduction is the $\delta \rho$ parameter that must lie below 0.4$\%$. The first concern is the two loop contribution to the W and Z masses from the exchange of a single diagonal ETC gauge boson across the techniquark loop contributing to $M_Z$ \cite{Chivukula:1995dc,Appelquist:1996kp}. Naively this gives a contribution
\begin{equation} \delta \rho =  {g^2 v^4 \over 12 \Lambda^2} \label{deltarho2loopod}\end{equation}
We plot the excluded range from this estimate in Fig \ref{gvsLam} as the shaded region. In fact the group theory coefficient of $1/12$ enables this bound to be evaded even for $\Lambda \simeq 3$TeV. The generic lesson though is that moving to larger cut off with a strengthening ETC coupling or moving to larger $N_f$ to enhance walking both move the model away from the excluded region.

Even in these cases which escape the first contribution to $\delta \rho$ there is a secondary contribution that can dominate \cite{Appelquist:1989ps}. The ETC interaction that breaks isospin strongly to generate the top bottom mass splitting can also enter into the techni-U and techni-D masses generating a large splitting there also. This we can calculate here explicitly and compare to the result in (\ref{deltarho3loop}). This equation places a bound of 100 GeV on the mass splitting. We will attempt a holographic non-perturbative computation to compare below.

\begin{figure}[]
\centering
\includegraphics[width=8cm]{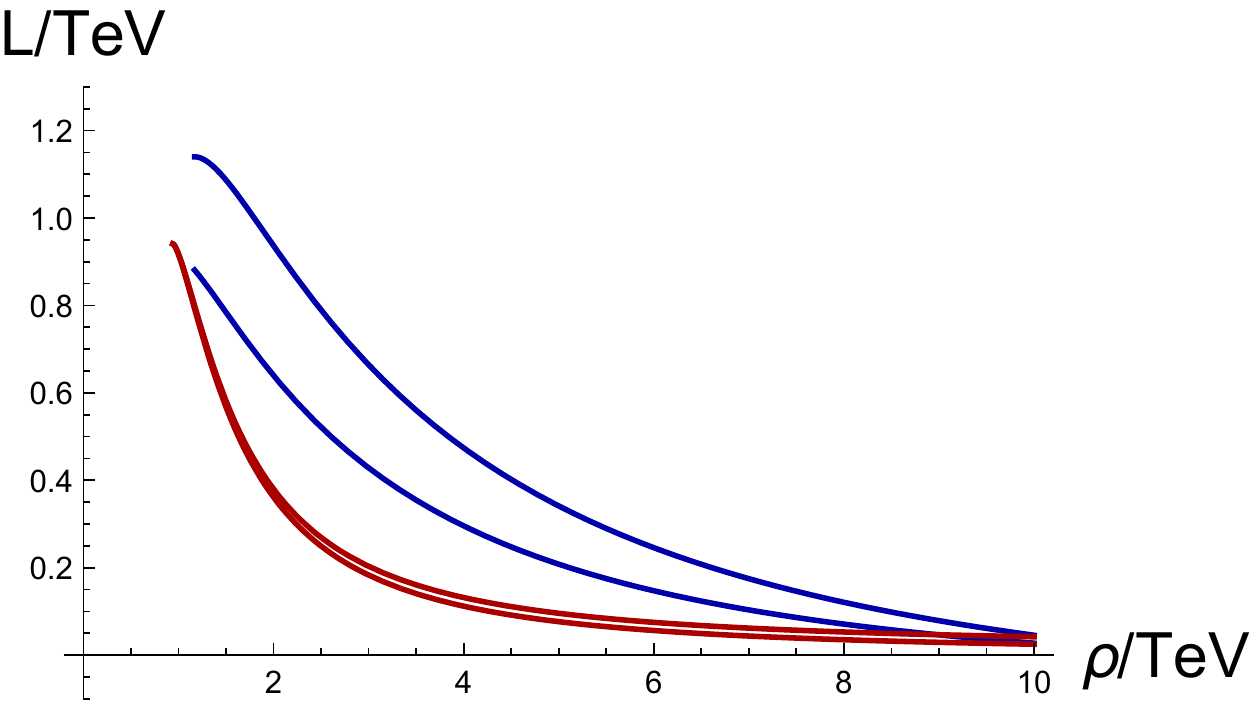}
\caption{ The self energy function $L(\rho)$ for the U (higher) and D (lower)  techniquarks against RG scale $\rho$ for solutions with the physical top mass, $N_c=3$, $N_f=2$ (lower two curves in the IR) and 11 (higher two curves in the IR). Here $\Lambda=10$ TeV }
\label{embed}
\end{figure}

Let us first plot a sample of the embedding functions $L(\rho)$ for the techniquarks, $U$ and $D$ - see Fig \ref{embed}. It is good intuition, for comparison to gap equation analysis, to treat these as the self energy function $\Sigma(p)$ for the quark. Broadly the IR is dominated by the TC dynamics and the self energies of the U and D are degenerate there whilst in the UV the four fermion interaction generates a UV mass splitting. The scales of these are set by $f_\pi$ and $m_t^{\rm phys}$ respectively. In Fig \ref{embed} we have shown examples for $N_f=2$ and $11$ to show there is  some $N_f$ dependence. In particular the walking theory where the TC coupling is stronger in the UV leads to the IR theory displaying more isospin breaking. 

\begin{figure}[]
\centering
\includegraphics[width=8cm]{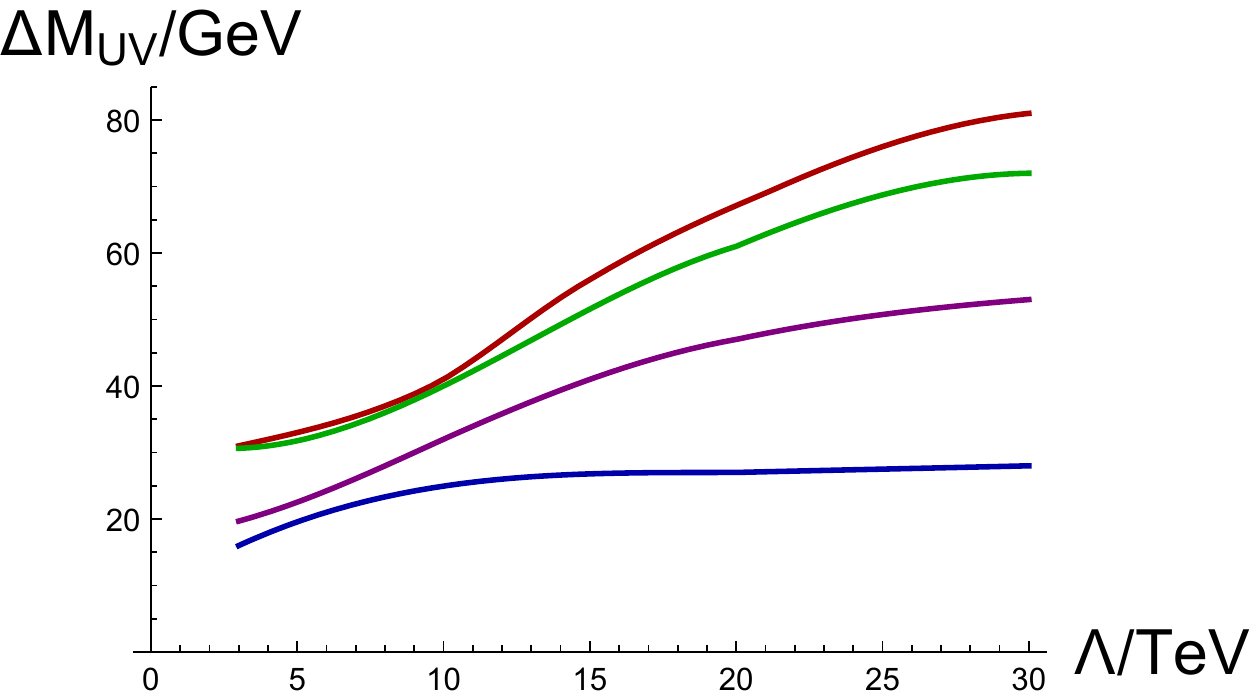}
\caption{ The UV cut off difference in the mass of the U and D techni-quarks for solutions with the physical top mass, $N_c=3$, $N_f=4,8,11$ from top to bottom.  }
\label{mUV}
\end{figure}

\begin{figure}[]
\centering
\includegraphics[width=8cm]{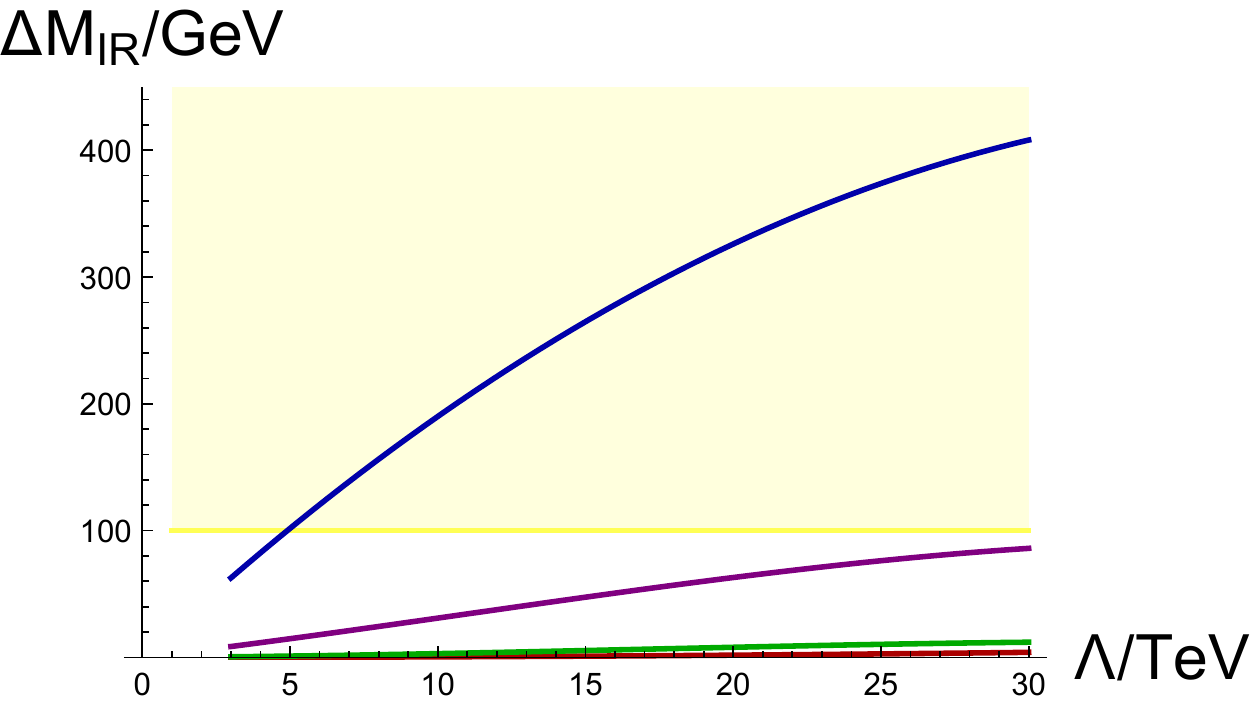}
\caption{ The difference in the mass of the U and D techni-quarks in the deep IR for solutions with the physical top mass, $N_c=3$, $N_f=4,8,11$ from bottom to top. The shaded region is excluded by the perturbative mass splitting computation of $\delta \rho$.  }
\label{mIR}
\end{figure}

To study this further we plot the mass splitting between the U and D in the UV as a function of $N_f$ and $\Lambda$ in Fig \ref{mUV}. The $N_f=2$ curve is at the top, $N_f=4,8$ central, and the $N_f=11$ curve is lower. This ordering reflects the growth of the condensate due to walking. The NJL interaction is weaker in walking theories to generate a given $m_t^{\rm phys}$.  Note all of these values lie below the 100 GeV naive bound.

In Fig \ref{mIR} we show the deep infra-red mass splitting between the U and D techniquarks for the solutions at each $\Lambda_{ETC}$ and for $N_f=2,4,8,11$. Here $N_f=2$ is the lower plot, $N_f=$ 11 the higher curve (the reverse of the UV behaviour). In the strongly running theories at low $N_f$ the symmetry breaking is dominated at low scales and the UV physics is suppressed since it lives in the asymptotically free regime of the theory - there is little IR mass splitting in the techni-sector. As we increase the anomalous dimension at the UV scale by walking we make the UV physics more important to the IR symmetry breaking and the NJL interaction plays a bigger role in enhancing the IR mass splitting. This model suggests that the gain of less splitting in the UV with walking is more than compensated by extra splitting in the IR. By $N_f=11$ the mass splitting in the techni-quark sector is greater than the 100 GeV perturbative bound (shown as the shaded area in the plot). 

Naively at this stage the $N_f\leq 8$ theories at a cut off scale up to 30 TeV appears to avoid all the $\delta \rho$ bounds: both that in Fig \ref{gvsLam} and  with the mass splitting in the technisector being below 100 GeV at all scales as shown Fig \ref{mUV} and Fig \ref{mIR}. 

\begin{figure}[]
\centering
\includegraphics[width=8cm]{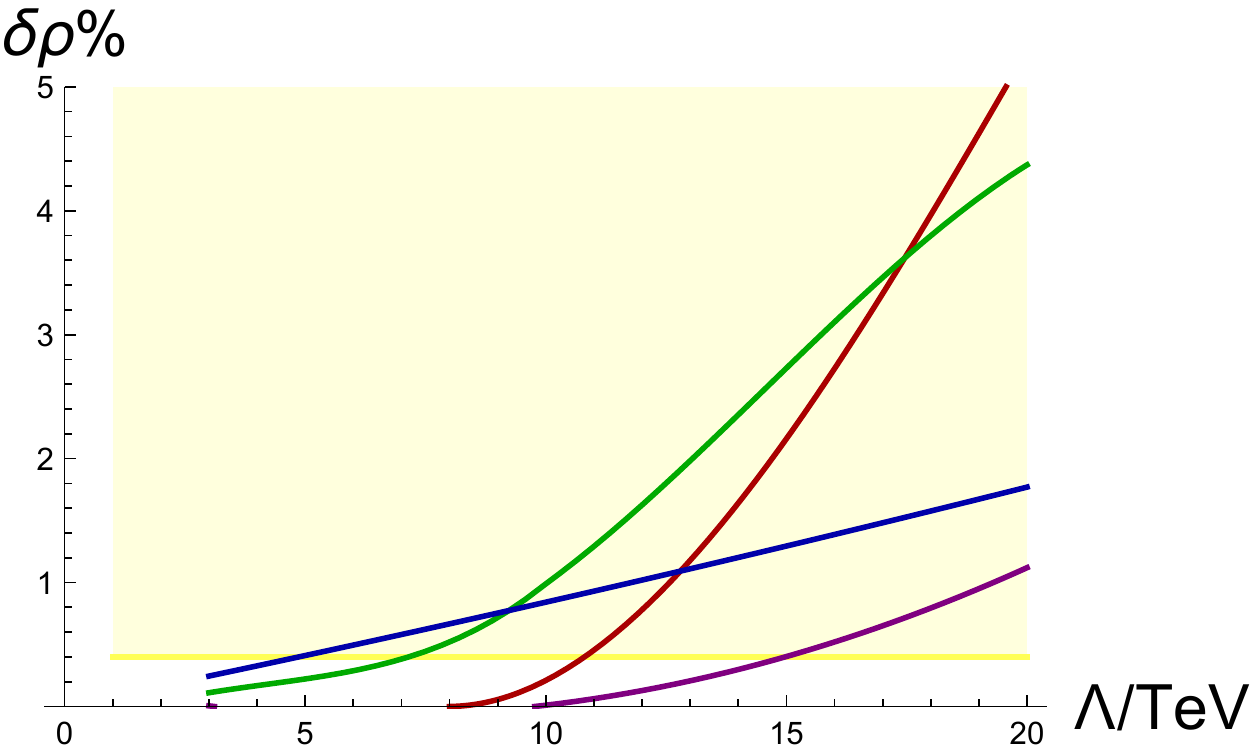}
\caption{ The holographic computation of $\delta \rho$ for solutions with the physical top mass, $N_c=3$. Moving down the right hand side of the plot are the curves for $N_f=2,4,11,8$ . The shaded region is experimentally excluded. }
\label{deltarho}
\end{figure}

Ideally one would like to compute the $\delta \rho$ contribution directly in our holographic model. Technically it is hard to compute $f_{\pi^{\pm}}$ holographically because in full string models $\bar{U} D$ states are described by true strings stretching between the U and D flavour branes. The spirit, as can be seen from the non-abelian Dirac Born Infeld action (which is only known for very small mass splittings) \cite{Myers:1999ps,Erdmenger:2007vj}, would be that the $f_{\pi^{\pm}}$ calculation would be some smearing over the two brane geometries. A reasonable proposal for this computation at the field theory level would be to replace (\ref{aa}) (for the pion)  with

\begin{equation}  \label{aa} \partial_\rho \left[ \rho^3 \partial_\rho A \right]   - 
\kappa^2 {{1 \over 4} (L_U + L_D)^2 \rho^3 \over  (L_U^2 + \rho^2)(L_D^2 + \rho^2)} A  = 0\,. \end{equation}

We then have
\begin{equation} \delta \rho = {f_{\pi 0}^2 - f_{\pi_\pm}^2 \over f_{\pi 0}^2} \end{equation}
We can compute this for the cases we have considered - the results for $N_f=2,4,8,11$ as a function of $\Lambda$ are shown in Fig \ref{deltarho} where it can be seen that the result is considerably larger than the perturbative estimate in (\ref{deltarho3loop}) suggests. The holographic computation of $f_\pi$ depends on more than just the magnitude of the self energy functions but also on derivatives etc (in this sense it is like the Pagel Stoker formula \cite{Pagels:1979hd} used with gap equations) and so can reasonably produce a larger result. There is also no clear pattern of behaviour with $N_f$ which is directly attributable to the fact that the IR mass splitting grows with $N_f$ whilst the UV splitting falls. Nevertheless the $N_f=8$ theory with a judicious amount of walking and moderately strong ETC appears able to survive constraints until a cut off of 15 TeV.

\section{Top Condensation Assisted Technicolour}

The difficulties of hiding the top mass generation mechanism from the $\delta \rho$ parameter are not new although our computational framework clearly presents them. Previously it has been suggested that the problem can be alleviated by an additional top self interaction as in top condensation models \cite{Miransky:1988xi,Miransky:1989ds,Nambu:1989jt,Bardeen:1989ds}. Clearly if a separate NJL model does the work of generating the top quark mass then the ETC interactions that feed that mass back into the techniquark masses will be reduced. Top colour \cite{Hill:1994hp} is an example of a model underlying such a mechanism.

\begin{figure}[]
\centering
\includegraphics[width=8cm]{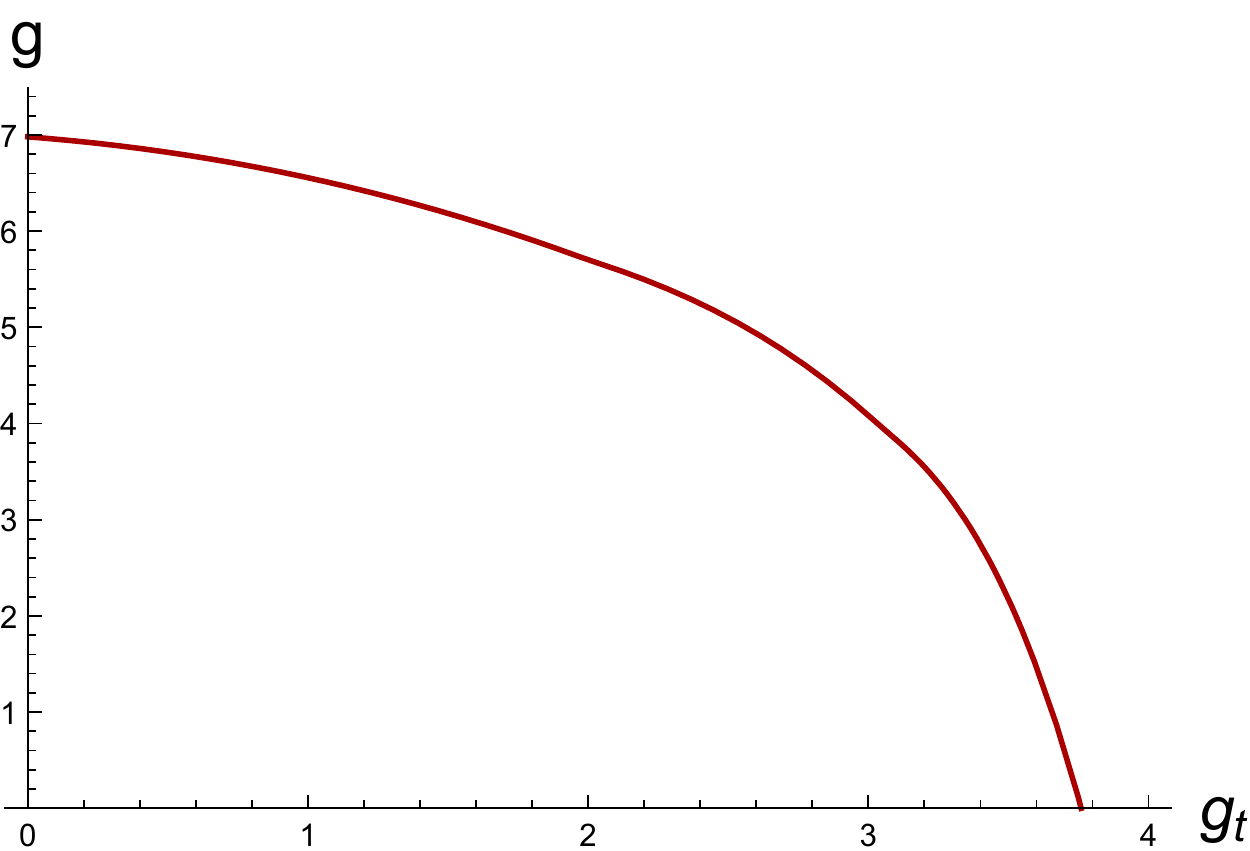}
\caption{ The top condensate coupling against the ETC coupling for solutions with the physical top mass, $N_c=3$, $N_f=2$ $\Lambda=5$ TeV.  }
\label{g2vsg}
\end{figure}

In the one doublet models we can just include a top self interation with coupling $g_t$ in (\ref{g2}) 

\begin{equation}  m_{t} = {g^2 \over 12 \Lambda_{ETC}^2} c_{t} + {g^2 \over 2 \Lambda_{ETC}^2} c_{U} + {g_t^2 \over \Lambda_{ETC}^2} c_t \label{g3} \end{equation}

For example we can compute with $N_f=2$ at $\Lambda=5$TeV, setting $m_t^{\rm phys}=175$ GeV in the IR. At each value of $\alpha_{TC}$ we tune the UV mass of the $U$ embedding to give the physical $f_\pi$ and then read off $g$ from (\ref{g1}). $g_t$ then follows from  (\ref{g3}). In Fig \ref{g2vsg} we plot the $g$ vs $g_t$ line that achieves the physical top mass and electroweak scale. As advertised one can trade the strength of the ETC interactions for a stronger top NJL coupling. In principle this can be used in any case to solve the $\delta \rho$ problem from the techni-sector.  

\section{A Hologram of One Family Technicolour}

\begin{figure}[]
\centering
\includegraphics[width=8cm]{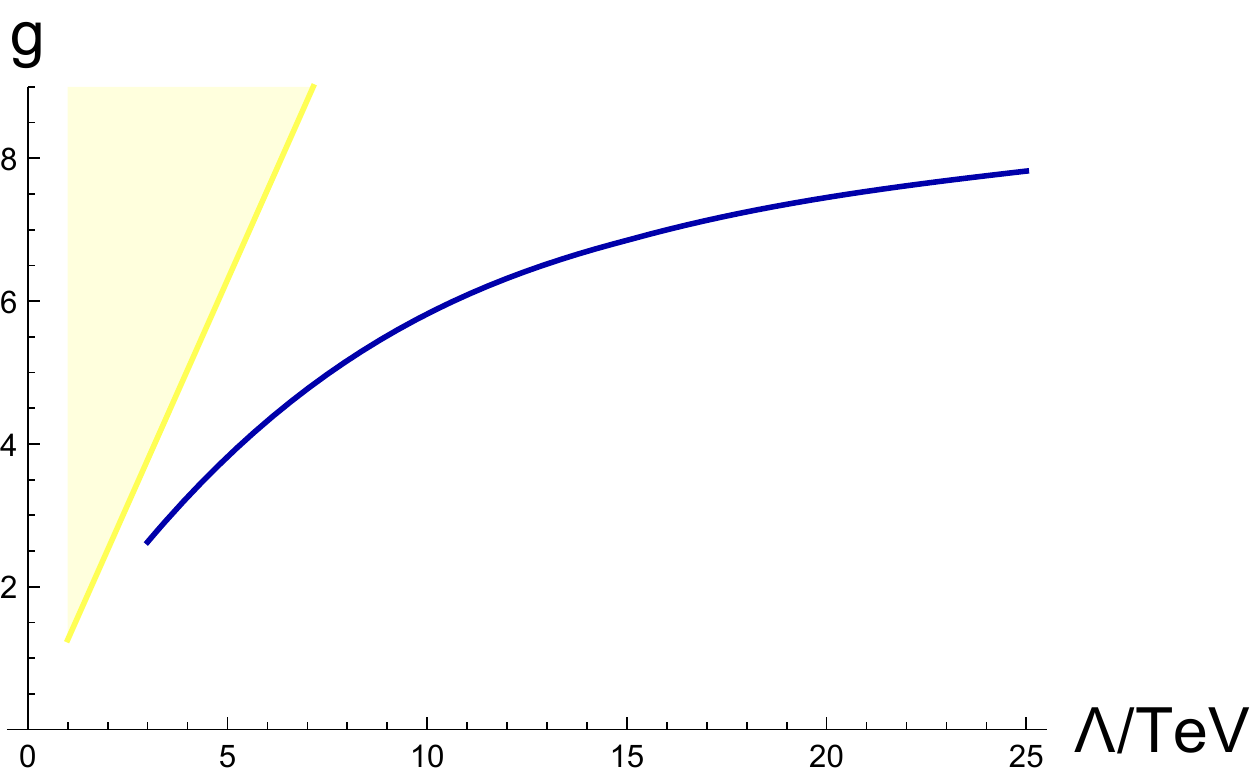}
\caption{ The ETC coupling against  ETC scale in the one family TC model.   }
\label{1fglam}
\end{figure}

Another classic ETC configuration is to have a full family of techni-fermions ($U^i$, $D^i$, E, N, ie $N_f=8$) each in the fundamental representation of SU($N_{TC}$). The minimal ETC group to generate just the top mass is to place $\Psi_L^i=(U^i,D^i)_L$ and $\psi_L^i = (t^i,b^i)_L$ and $t_R^i$ in the fundamental representation  of an SU($N_{TC}+1$) ETC group that is then broken at the scale $\Lambda$ to the technicolour group. The broken step generators lead to the four fermion operators
\begin{equation} {g^2 \over 2} \bar{\Psi}_L^i t_R^i \bar{U}_R^i \psi_L^i \end{equation}
and the diagonal generator for $N_{TC}$=3 ($1/\sqrt{24}$ diag(1,1,1,-3) ) gives
\begin{equation} {g^2 \over 24} \bar{\Psi}_L^i U_R^i \bar{U}_R^i \Psi_L^i + {9 g^2 \over 24} \bar{\psi}_L^i t_R^i \bar{t}_R^i \psi_L^i\end{equation}
where here the colour index $i$ is not summed over.

The holographic description is as follows. The QCD sector is described by Dynamic AdS/QCD with $N_c=3$ and $N_f=6$ - the model predicts the quark condensate at the ETC scale which we must divide by 3 to get the condensate contribution from a single QCD colour of quark. In principle one ought to adjust the QCD running above the techni-quark mass, however, since this is in the slow running perturbative regime for the QCD coupling where the top quark mass runs very slowly we neglect this complication. 

The technicolour sector is described by a Dynamic AdS/QCD model with $N_{TC}=3$ and $N_f=8$. As before we require the $m_D=m_E=m_N=0$ at the ETC scale. These five fermions contribute degenerately to $f_\pi$. We can then dial $\alpha_{TC}$($e^2$TeV) and $m_U^{IR}$ to generate configurations with the correct electroweak $f_\pi=246$GeV - the three colours of techni-Us and the top also contribute here. To determine the correct combination of top and techni-up embeddings we now require at the UV scale that 
 
\begin{equation}  m_{U} = {g^2  \over 24 \Lambda^2} c_{U} + {g^2 \over 2 \Lambda^2} {c_{t} \over N_c}\end{equation}
and 
\begin{equation}  m_{t} = {9 g^2 \over 24 \Lambda^2} {c_{t} \over N_c} + {g^2 \over 2 \Lambda^2} c_{U}  \end{equation}

\begin{figure}[]
\centering
\includegraphics[width=8cm]{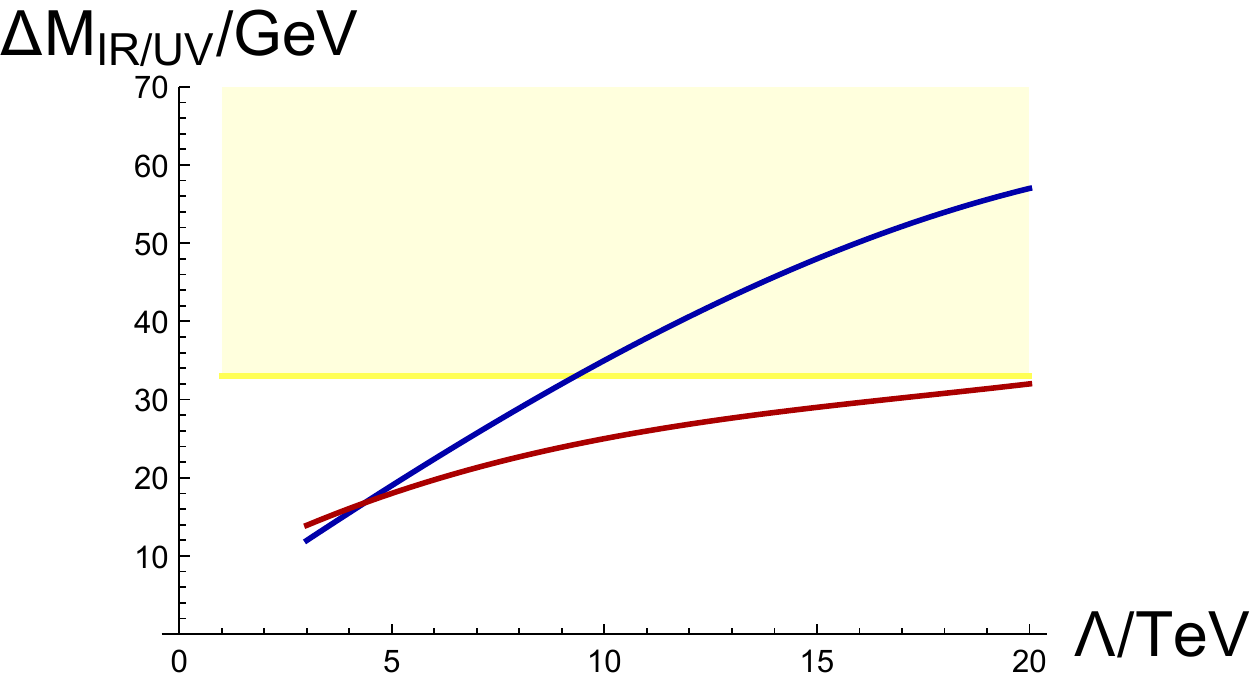}
\caption{U-D mass splittings in the one family TC model against $\Lambda$. On the right of the plot, the top line is the IR mass splitting, the lower line the UV splitting.    }
\label{1fdelm}
\end{figure}

\begin{figure}[]
\centering
\includegraphics[width=8cm]{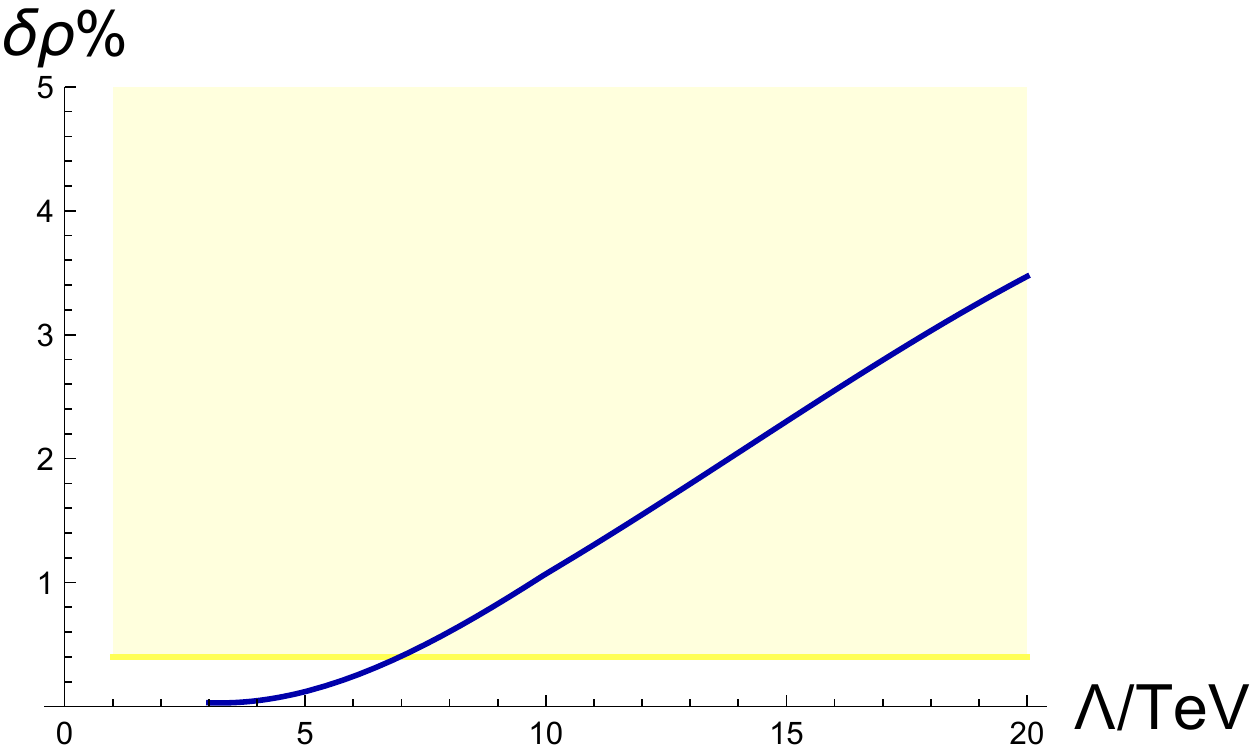}
\caption{$\delta \rho$ in the one family model as a function of ETC scale.  The yellow region is excluded by the experimental bound.   }
\label{1fdelrho}
\end{figure}

We again plot $g$ vs $\Lambda$ for the model in Fig \ref{1fglam} where we see that strong ETC values of $g$ are required to generate $m_t^{\rm phys} =175$ GeV. There is again a second NJL dominated branch of solutions which we don't show - as in the one doublet model these have very large isospin breaking in the techni-quark doublet.  We plot the mass splittings in the techni-quark doublets $\Delta m_{IR}$ (evaluated in the IR) and $\Delta m_{UV}$ (evaluated at the ETC scale) against $\Lambda$ for the technicolour dominated solutions in Fig \ref{1fdelm}. The splittings are between 10 and 70 GeV. Remember here that there are 3 electroweak doublets with this splitting contributing to $\delta \rho$ so much of the range is excluded again even by the pertubative estimate of $\delta \rho$ - the excluded region is shaded in the plot.  Finally we plot the values of $\delta \rho$ from the holographic computation of $f_{\pi^\pm}$ which are considerably larger and exclude the model for ETC scales above 7 TeV. 

In conclusion the one family model struggles more on all phenomenological fronts from S to $\delta \rho= \alpha T$! Of course a direct top condensation NJL interaction could again be used to decouple the techniquark sector from the isospin breaking of the top mass.

\section{Discussion}

Holographic models provide a  calculationally efficient tool to study the broad behaviours of strongly coupled gauge theories. They incorporate the ideas of walking dynamics very directly since the AdS mass of states translates to the running anomalous dimension of the quark condensate, $\gamma$. The Dynamic AdS/QCD model we have used here is a very simple crystallization of these ideas inspired by top-down string models. It allows the study of the mesonic sector of any theory if a sensible guess is made for the running of $\gamma$ - here we have used the two loop running of the gauge coupling which incorporate the physics of the conformal window, chiral symmetry breaking when $\gamma=1$, walking for theories just above that point in $N_f$ and then QCD-like dynamics for smaller $N_f$.

Four fermion NJL operators can be included using Witten's multi-trace prescription and the critical behaviour of the NJL model can be realized. Here we have included NJL operators in Dynamic AdS/QCD and again shown traditionally NJL like behaviour (see also \cite{Clemens:2017udk}). We have designed descriptions of  dynamical symmetry breaking models of the electroweak sector. A pure NJL model can be used to generate a top quark condensate - in Fig \ref{topmass} we show the rise of the top mass above some critical NJL coupling (close to the perturbative $2 \pi$ value for the coupling). 

We have then studied the interplay between a strongly coupled technicolour gauge theory  and the top quark, linking the two sectors by extended technicolour NJL operators. In our first model we used NJL operators inspired by a one doublet technicolour model (with $N_c=3$) and the simplest extended technicolour unification of the top quark. We allowed for possible electroweak singlet techniquarks to vary the total $N_f$ of the gauge theory. 
We studied solutions where the electroweak $f_\pi$ is achieved and various values of the top mass. There are two possible solutions. One is an NJL dominated solution where the four fermion operators drive the majority of the electroweak breaking and technicolour is relatively weak; these models have large isospin breaking in the technicolour sector and are ruled out by precision data for $\delta \rho$. The second set of solutions match to traditional technicolour dominated electroweak symmetry breaking and weak ETC for small top masses. The dynamics can be followed here to strong ETC couplings and large top masses. A maximum top mass is possible without creating too large an $f_\pi$. For theories at low $N_f$ this top mass value is close to 500 GeV but it increases significantly as $N_f$ approaches the edge of the conformal window and walking enhances the techniquark condensate. When we studied solutions with the physical value of the top mass the holographic model allows us to study the IR and UV mass splittings in the techniquark sector induced by the isospin violating ETC interactions. These splittings lie between 20 and a few hundred GeV and taken naively with the perturbative expression for $\delta \rho$ (\ref{deltarho3loop}) suggest models may be compatible with electroweak data. However, we also used the holographic model to estimate $f_{\pi^\pm}$ and directly determine $\delta \rho$ and these estimates were up to an order of magnitude larger (due to isospin violating structure in the derivatives of the techniquark self energy)   ruling out larger ETC scales. A judicious choice of ETC scale near 3-15 TeV, a modicum of walking (too much enhances the isospin breaking effects of the UV ETC interactions) and strong ETC should pass the experimental bounds though. In conclusion the holographic model provides a clean computational framework that emphasises the roles of walking dynamics and strong ETC interactions in the top mass ETC generation mechanism.  A separate NJL interaction to generate the top mass can be used (as in top colour models) to isolate the isospin breaking of the top mass from the technicolour sector. 

Finally we studied a one family technicolour model with $N_c=3$ and $N_f=8$ and observed the same structure of solutions. Here because of the three QCD colours of techniquarks the isospin splitting in the techniquark sector makes a larger contribution to $\delta \rho$ and these models are harder to reconcile with experiment although an ETC scale between 3-7 TeV seems possible.

Whilst many of the phenomenological conclusions of this analysis have been previously intuited in other ways we believe that the holographic approach to the problem provides a simple and revealing computational tool that has made it worth studying independently of the precise phenomenology. We hope  that holographic models can play an important part in understanding strongly coupled sectors of beyond the standard model sectors in the future. 

\bigskip \bigskip

\noindent{\bf Acknowledgements:} NE's work was supported by the STFC consolidated grant ST/L000296/1 and WC's by an STFC studentship.

\end{document}